\documentclass[journal=jacsat,manuscript=article]{achemso}

\usepackage[version=3]{mhchem} 
\usepackage{tabularx}

\author{Sander Roet}
\affiliation[NTNU]{Department of Chemistry, Norwegian University of Science and Technology, Trondheim, Norway}
\email{sander.roet@ntnu.no} 

\author{Christopher David Daub}
\affiliation[UofH]{Department of Chemistry, University of Helsinki, P.O. Box 55, FI-00014, Helsinki, Finland}

\author{Enrico Riccardi}
\affiliation[UiO]{Department of Informatics, UiO, Gaustadall\'{e}en 23B, 0373 Oslo, Norway}

\title[Chemistrees]
  {Chemistrees: \\ data driven identification of reaction pathways via machine learning.}

 
\begin{document}
 


 
\begin{abstract}

We propose a supervised machine learning algorithm, decision trees, to analyze molecular dynamics output. The approach aims to identify the predominant geometric features which correlate with trajectories that transition between two arbitrarily defined states. The data-based algorithm aims to identify such features in an approach which is unbiased by human "chemical intuition".

We demonstrate the method by analysing proton exchange reactions in formic acid (FA) solvated in small water clusters. The simulations were performed with \emph{ab initio} molecular dynamics combined with a method for generating rare events, specifically path sampling. Our machine learning analysis identified mechanistic descriptions of the proton transfer reaction for the different water clusters. 

\end{abstract}

\section{Introduction}
In regions far from urban areas, formic acid (FA) has been recognised as one of the main factors which reduces the pH of rainwater, causing acid rain.~\cite{Murdachaew2016a} It has relatively high atmospheric concentrations \cite{Millet2015a,Chaliyakunnel2016a}, and contributes to the formation of sulphuric acid in the atmosphere~\cite{Kangas2020a,Daub2020a}. Enhanced description and prediction of proton exchange reactions involving solvated FA can improve current atmospheric models.

\emph{Ab initio} molecular dynamics simulations have recently been used to examine FA deprotonation in aqueous solution~\cite{Murdachaew2016a, Daub2019b}, demonstrating that simulations can successfully model proton exchange between water and FA.  While these studies led to valuable new insights, the limitations of the adopted methods (\emph{e.g.} usage of a bias potential, continuous collective variables) could be overcome thanks to relatively novel methodologies, such as Replica Exchange Transition Interface Sampling (RETIS)~\cite{vanErp2007a, Riccardi2017a}. Respecting the natural dynamics of the system allows the study of transitions even with a significant diffusive contribution \cite{riccardi2019predicting, riccardi2020permeation} (\emph{i.e.} a small reaction barrier), and enables the direct investigation of reaction mechanisms.

RETIS is a rare event method developed to investigate transitions. Its main advantages are: (a) it does not alter the natural dynamics of the system, (b) it does not require a particularly accurate order parameter, (c) its results are in principle identical to what would be obtained by an infinitely long unbiased molecular dynamics simulations. With RETIS, the transition region is explored by continuously generating new paths which start from a stable state and end up either back in such a state -- an \emph{unreactive} path, or reach a different state  -- a \emph{reactive} path. The approach has been successfully employed to study transitions that would, otherwise, require prohibitively long simulation times. The results generated have been used to describe the dynamics of chemical processes (\emph{e.g.} reaction rates) while considering the entropic contribution in the analysis\cite{moqadam2017rare, Moqadam2018a, Daub2019b, riccardi2019predicting, riccardi2020permeation}. Since large quantities of data are often generated by the sampling procedure,  approaches to pragmatically decode reaction mechanisms are greatly beneficial. 

Our aim is to establish an heuristic approach to describe transitions, regardless of whether they involve crossing an entropic barrier. Data driven, physically consistent and measurable system descriptors might be generated, and their correlations with the system dynamics asserted. This is a classification problem which a Machine Learning (ML) algorithm can be trained to solve. The algorithm might then predict if a certain molecular structure (frame) is part of a reactive or a non-reactive trajectory. 
Connecting the descriptors to measurable quantities provides a data driven ("unbiased") description of a transition that might support, and eventually surpass, human-biased "chemical intuition". 

Data driven algorithms for enhanced sampling or the analysis of chemical simulations have significant recent contributions.~\cite{van2016analyzing, Hooft2021, Chen2018, Schberl2019, Ribeiro2018, jung2019}. Most of these approaches are based on neural networks, which lack physically consistent interpretability. By contrast, this can be a desirable feature of decision trees~\cite{Chen2018, Schberl2019}. The usage of neural networks in rare events, furthermore, requires a pre-selection of trial collective variables~\cite{Hooft2021, Ribeiro2018, jung2019}, which could lead to a hypothesis bias. 
Decision tree \cite{swain1977decision} (DT) classifiers have an unique solution and are not sensitive to highly correlated variables. The results can be readily interpreted if the source variables are also interpretable.  This approach was previously adopted to select optimal collective variables with decision trees, with reasonable success.~\cite{Moqadam2018a}

Our work based on decision trees provides both an interpretable and hypothesis-bias-free method and proposes the usage of interpretable DTs via an  appropriate system representation invariant to system translation, rotation, and changes in atomic indices. Our aim is to gain insights into reaction mechanisms with a systematic and unbiased representation of the system.  

The approach has been developed with sufficient versatility to be applied to different types of molecular simulations, from conventional molecular dynamics (MD) to rare event methods. It shall be noted that conventional MD would require \emph{a priori} classification of the data (\emph{i.e.} dividing the source trajectory into reactive and unreactive segments). The sampling strategy of rare event methods, instead, generates a data structure which inherently classifies the trajectories. Regardless of the adopted molecular simulation approach, limiting the correlation between samples is a primary task for a quantitative data driven method to identify reaction paths and the probability of their occurrence.

In the present study, we focus 
on small clusters of formic acid solvated by water, HCOOH + (H$_2$O)$_n$, $n=4$ and $6$. The system is relatively small and well understood, and hence provides an ideal test case for training a ML method.  At the same time, studying the system using our  methodology provides new quantitative and qualitative insights into the acid-water proton transfer reaction which is also of interest in atmospheric chemistry.~\cite{Leopold2011a,Forbert2011a,Chung2017a,Lengyel2017a,Gutberlet2009a,Maity2013a,Elena2013a}.

\section{Computational Models and Methods}

Our main focus is the machine learning methodology here introduced, and therefore only a brief introduction to the simulation methodology will be provided. Please consider our previous work\cite{Daub2019b,Daub2020a} for further details.

\subsection{System description}

For studying proton transport, molecular simulations able to consider bond formation and bond breaking are required. Born-Oppenheimer molecular dynamics (BOMD) has been shown to be a suitable approximation in previous studies of atmospheric reactions~\cite{Kangas2020a,Daub2020a} and of aqueous formic acid~\cite{Murdachaew2016a,Daub2019b}. The density functional theory BLYP implemented in the Quickstep module of CP2K~\cite{VanDeVondele2005a} has been combined with a double-zeta basis set supplemented by the use of Grimme's D2 dispersion correction.~\cite{grimme2010consistent} 

\begin{figure}
  \includegraphics[width=0.3\textwidth,trim={0.2cm 0.2cm 22cm 0},clip]{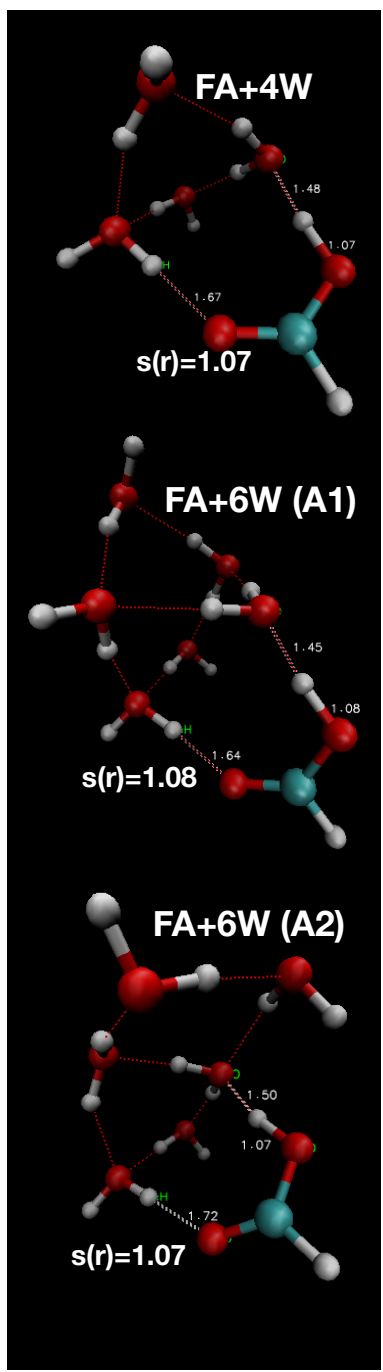}
  \caption{Minimum energy configurations for systems with formic acid (FA) associated with 4 and 6 water molecules.  $s(r)$ in all cases shown here is equivalent to the $r_{OH}$ of the initially protonated formic acid (FA) molecule.  These configurations correspond to the initial states used to start PyRETIS simulations.}
  \label{fig:inits}
\end{figure}

A set of systems with an increasing number of water molecules around FA were studied. Initial configurations were obtained from minimum energy configurations, of which snapshots are reported in Figure~\ref{fig:inits}. As more water molecules were added, the probability of generating reactive trajectories  
became larger.  However, at least four added water molecules were required to allow generation of trajectories with a significant charge separation between the deprotonated FA and the solvated proton. Adding two more water molecules resulted in a significantly higher proton transfer rate.
The two systems composed by the formic acid surrounded by four and six water molecules are here analysed and discussed.

\subsection{Definition of the Collective Variable}

Path sampling simulations require the definition of a collective variable, $s(r)$, where $r$ represents the set of positions and velocities of all atoms in the system. The collective variable must describe the transition from a reactant state A to a product state B.  Path sampling is not limited to continuous variables, which  
allows the consideration of relatively more complex functions to describe  
proton transport.

The collective variable here adopted, $s(r)$,  
is inspired by the study of water ionization~\cite{Moqadam2018a}, with modifications introduced to consider acid deprotonation.  As a first step, it locates the smallest distance between any formic acid oxygen and any reactive hydrogen in the system excluding the methyl hydrogen in formic acid.  This distance is denoted as $r_{O_{FA}H,min}$.
 
For $r_{O_{FA}H,min} < 1.4$ \AA\, the formic acid can be considered protonated, so
$s(r) = r_{O_{FA}H,min}$.  For $r_{O_{FA}H,min} > 1.4$ \AA, charge separation between the solvated proton and the formic acid becomes significant. To quantify it, and thus compute $s(r)$, all the distances between reactive hydrogens and oxygens are first considered. Hydrogens are then assigned to the closest oxygen, either water or formic acid.  Any water oxygen found to be associated with 3 hydrogens is indexed.  Next, all distances between formic acid oxygens and hydrogens associated with triply coordinated water oxygens are sorted. $s(r)$ is the minimum value of these distances.

Conceptually, the reasoning behind this complex definition of $s(r)$ is to allow the description of formation of more extended systems such as Eigen or Zundel cations.  A discontinuous jump of $s(r)$ from $\sim 1.8$ \AA\ up to $\sim 3$ \AA\ is associated with a change in the identity of the triply coordinated water oxygen and the formation of structures resembling Zundel cations H$_5$O$_2^+$.  The formation of the Zundel cation with $s(r) > 2.9$ \AA\ is here labeled as the product state B.

\subsection{RETIS}
The PyRETIS \cite{lervik2017pyretis, riccardi2020pyretis} library has been used to perform RETIS\cite{vanErp2003a} simulation, coupled with the \textit{ab-initio} molecular dynamics external engine, CP2K. The initial paths describing the transition from protonated to deprotonated FA along $s(r)$ were generated by using the \emph{kick} method available in the software, starting from the initial configurations shown in Figure~\ref{fig:inits}. The \emph{kick} approach uses a mixture of stochastic and deterministic dynamics to generate a set of initial paths. The results were obtained by discarding paths correlated with the initial generated ones. Finally, the remaining paths from three and four independent simulations for the four and six water cases , respectively, were merged together.

The RETIS simulations varied somewhat in the details of the number of interfaces used for the transition interface sampling, and the location of the first interface (\emph{i.e.} the interface between the $[0^-]$ and the $[0^+]$ ensemble).  In the 4 water simulations, the first interface was placed at $s(r) = 1.05$ \AA, and the last interface at $s(r) = 3.0$ \AA.  
Similarly, the 
6 water simulations had the first interface at $s(r) = 1.07$ \AA, and the last interface at $s(r) = 3.0$ \AA.  

\subsection{Selection window}

In a trajectory, each frame can be considered as an instance. Depending on the simulations set up, a large number of frames would generate a long list of instances with a very high correlation. Furthermore, different trajectories can be highly correlated one another, depending on the sampling algorithm. 
Since generating a sufficient number of uncorrelated trajectories often requires excessive computational requirements, an approach to provide a sufficient sampling with a limited correlation is here proposed. 
We have thus selected frames contained in a rather restricted region in path space, the selection windows. By randomly picking a certain number of frames, for each trajectory, within a selection window, the correlation between instances is minimized. Depending on the location and size of the selection window, different outcomes can be expected. By placing it, as in the current study, in proximity of the initial state, the system configurations that are correlated with the transitions can be identified at an early stage of the transition.
The selection window location and dimension variables, along with the number of frames per trajectory to consider, constitute the three hyper-parameters of our approach.
In the present work, the machine learning algorithm has been thus fed with one frame per trajectory with order parameters $1.1 < s(r) < 1.25$ \AA. The range is sufficiently narrow to consider only a few frames for each trajectory, each with similar order parameter.

The selected frames, each of them belonging to a different trajectory, have comparable values of the order parameter. The machine learning algorithm should, hereby, determine the most relevant feature(s) for the transition to happen, without bias on the main descriptor of the transition itself. This ensures that the detected features are not correlated to the classification of the trajectory.

\subsection{Training of the Decision trees}

Decision trees report the most important features to differentiate between reactive and unreactive paths, without imposing any prior hypothesis. In the present work, we considered atomic distances and velocities as possible features. Both entries have to be treated such to be translation and rotational invariant. As we did not achieve any significant improvement in the decision tree by considering atomic velocities, we here report the work based on the analysis of atomic distance only. 

In most of the simulation packages, Molecular simulations produce an ordered data array for the positions and velocities for each atom in a time frame. While the convention facilitates post processing and visualization procedures, it includes a bias in the data representation. Small deviations in the observation angle, or on the axis definition (e.g. exchanging $x$ with $y$ coordinates) leads to significantly different data sets, while corresponding to nearly identical systems. Data thus has to be pre-processed to become invariant with respect to translations and rotations. Furthermore,  the machine learning problem also has to be atom-index invariant, and the sorting method also must be reversible to allow back-mapping of the machine learning prediction  to the relevant atom (or atom pairs).

\begin{figure}
    \begin{flushleft}
    \begin{tabular}{l|rrrrr|}
        & C0 & H0 & O0 & O1 & H1\\
        \hline
        C0 & 0.0000 & 0.1109 & 0.1250 & 0.1321 & 0.2010 \\
        H0 & 0.1109 & 0.0000 & 0.2047 & 0.2013 & 0.2954 \\
        O0 & 0.1250 & 0.2047 & 0.0000 & 0.2311 & 0.2524 \\
        O1 & 0.1321 & 0.2013 & 0.2311 & 0.0000 & 0.1073 \\
        H1 & 0.2010 & 0.2954 & 0.2524 & 0.1073 & 0.0000 \\
        \hline
    \end{tabular}
    \end{flushleft}
    \hfill 
    \includegraphics[width=0.3\linewidth, clip, trim=55cm 15cm 55cm 14cm]{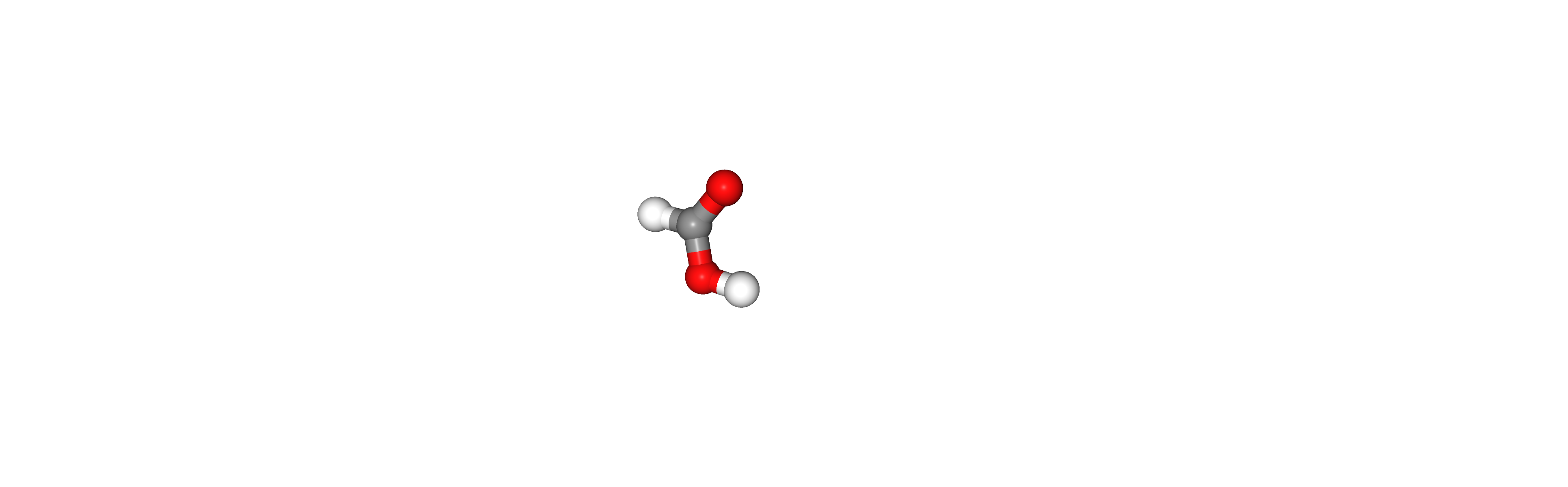}
    \begin{flushleft}
    \begin{tabular}{l|rrrrr|}
        & C0 & H0 & H1 & O0 & O1 \\
        \hline
        C0 & 0.0000 & 0.1109 & 0.2010 & 0.1250 & 0.1321 \\
        H0 & 0.1109 & 0.0000 & 0.2954 & 0.2013 & 0.2047 \\
        H1 & 0.2010 & 0.0000 & 0.2954 & 0.1073 & 0.2524 \\
        O0 & 0.1250 & 0.2047 & 0.2524 & 0.0000 & 0.2311  \\
        O1 & 0.1321 & 0.1073 & 0.2013 & 0.0000 & 0.2311 \\
        \hline
    \end{tabular}
    \end{flushleft} 
    \caption{Distance matrix (top) for a structure of formic acid (right). 
    The index invariant distance matrix (bottom) corresponding to the first distance matrix.} 
    \label{fig:distsolution}
\end{figure}

The translation and rotation independent requisites are met by a simple atom-atom distance matrix.
For the FA acid in a water cluster the carbon atom is the trivial identifiable reference atom. The atom-index invariant requisite is, on the other hand, a more elaborate representation. The matrix is created by first choosing an anchor point (C0) and grouping the row per element and sorting the distance with the anchor atom for each element. The resulting matrix reports the distance from a selected anchor atom (rows) to its next neighbour (columns). A schematic of the algorithm to generate both the simple distance matrix and the atom invariant distance matrix is provided in the SI.
In Figure \ref{fig:distsolution}, the two matrix representations are provided, as an example, for a formic acid molecule.
 
The resulting internal coordinate representation allows an independent analysis of each entry and, thus, a suitable data structure for the machine learning task.

Another common translational and rotational invariant representation, the Z-matrix \cite{baker1991geometry}, also provides an appealing internal representation of molecular structures as it scales better with the number of atoms compared to the distance matrix. Yet, the value of its variables (\emph{i.e.} distances, angles and dihedrals) are dependent on each other and on the atom sequence. In contrast, in our distance matrix representation, the considered variables are all independent. Also, distances are unique, with a lower (0) bound and an upper bound (system size). These three characteristics allow for a proper split of sample space to be made by the decision trees. Furthermore, our representation can be index invariant.

The machine learning problem we are posing is: what are the main features that a simulation frame has to have in order to be part of a trajectory that connects an initial state to a final state (reactive)? With which probability?
 
The information gain (entropy) decision tree is a viable method for a problem with highly correlated features.~\cite{Breiman2017}

Given a set of trajectories, a classification between reactive and unreactive paths is first needed. A numerical descriptor, conventionally defined as order parameter, can quantify the progress of a given transition. If its value, for a given system, is within certain arbitrarily defined ranges, the system can be considered to be located in the initial or final state.
A reactive path is defined as a path starting from an initial state and ending at a product state. A non reactive path, instead, ends at the initial state.
If the input generated by molecular simulation is composed by a single long trajectory, sub-segments will have to be fed to the machine learning task. If a segment starting at one state and ending in another state will be considered reactive, while a segment starting and ending at the same state without having entered another state before, is considered as unreactive.
When using the input generated by path sampling, paths contained in a single ensemble should be considered (please consider Refs. \citenum{vanErp2007a, Cabriolu2017a} for the definition of an ensemble and further details of the path sampling methods).
Regardless of the approach to generate input paths, or segments, it is worth noting here that the population of reactive and unreactive paths has to be sufficient to feed a machine learning algorithm. The necessary population size depends largely on the system size and type. In the attempt to provide an estimate of the reliability of the predictions, an error-estimate procedure has been also developed and described in a forthcoming section.

Computationally, a DecisionTree Classifier from scikit-learn~\cite{scikit-learn} has been fed with the index invariant matrix, using the 'entropy' splitting criterion and a max depth of 3.

\subsection{Visualizing decision paths}
 
A symmetric distance matrix can be back-mapped to $xyz$ coordinates (up to a translation and rotation), as described by \citet{Young1938} (and further detailed in the SI). The index-invariant distance matrix can be unsorted into the symmetric matrix, up to an atom index difference. The approach permits
to add dummy atoms according to the splits given by the decision tree, allowing a direct visualization of the analysis output (e.g. via VMD~\cite{Humphrey1996a}). For a convenient visualization, only the dummy atoms corresponding to the nodes along each decision path in the tree are selected. A main decision path is chosen such that the leaf node would have the highest number of pertinent reactive paths, weighted by the percentage of pertinent reactive paths: $n_r \cdot \frac{n_r}{n_r + n_u}$, where $n_r$ is the number of reactive paths in that node and $n_u$ the number of unreactive paths.

\subsection{Random forest decision error-estimate}
With highly correlated data, significantly different trees can be originated depending upon the first split from minor variations of the input. This is a constitutive limitation of the approach. The relative importance of the first split can be asserted by using Random Forests with a unit depth. The feature importance of such Random Forests is equal to the importance of the first split only. IT shall be noted that the feature that has the highest importance in the Random forest plot does not necessarily represent the main split for all possible decision trees.  

A sequence of forests of decision trees have been generated in respect with the time sequence on which sampling output has been generated. Source data has been split in 10 sub-blocks and randomized within. Random forest has been then computed for each of these subsets, generating a sort of time-dependent profile for the main splits, which allows to compute a variance $\sigma$ on each of the main features. The average value, for each feature, can be instead computed by considering the whole data set.
By assuming a Gaussian distribution for each feature, and by using the previously obtained variance and mean value for each feature, the relative probability of a feature importance can be estimated. By comparing the probability distribution for each feature, the most relevant can be identified even in the presence of highly correlated data.

Computationally, a RandomForest Classifier from scikit-learn~\cite{scikit-learn} has been fed with the index invariant matrix, using the 'entropy' splitting criterion and a max depth of 1.

\section{Results and discussion}

\begin{figure}
  \includegraphics[width=0.6\textwidth]{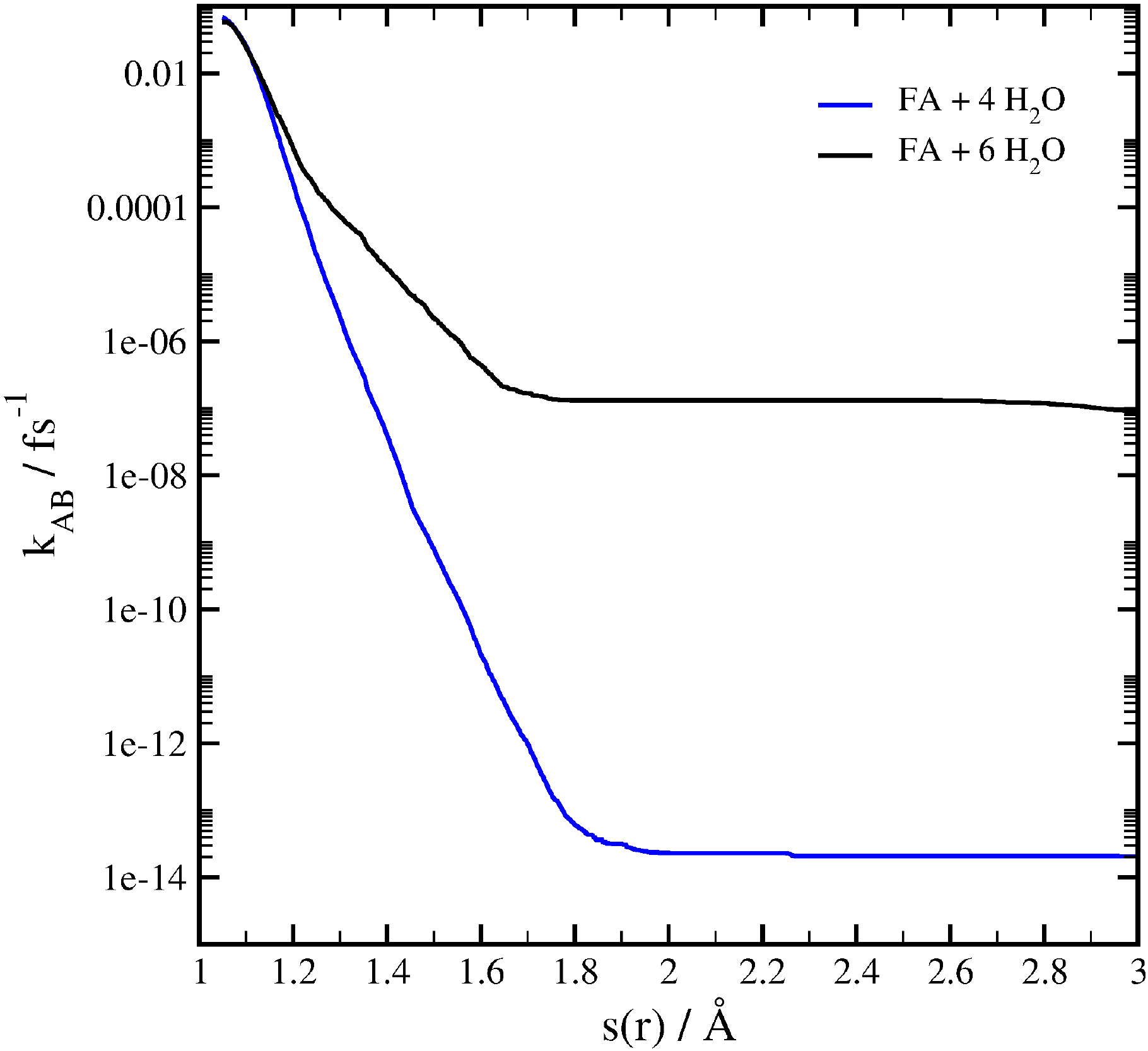}
  \caption{Effective rate constant $k_{AB}$ computed using RETIS for FA clustered with 4 or 6 molecules to reach the deprotonated state. Results are an average of several RETIS simulations from different initial conditions, weighted by the number of RETIS cycles in each simulation.}
  \label{fig:kload}
\end{figure}

The rate of proton transfer from the formic acid to the water molecules has been computed via RETIS simulation and ab-initio molecular dynamics simulations. Figure \ref{fig:kload} reports the rate of reaction for two systems, where 4 and 6 water molecules surrounded the FA acid. State A (protonated state) is defined as configurations with $s(r) < 1.05$ \AA (4 waters) or with $s(r) < 1.07$ \AA (6 waters).  State B includes  configurations with $s(r) > 3.0$ \AA for both systems.

The rate of proton transfer for the four water molecule case is $~ 2.10 10^-14 1/fs$, while $~ 1.01 10^-7 1/fs$ for six water molecules around the formic acid ($~ 10^7$ times difference).

We here investigate the mechanism of reactions via decision trees to identify the feature(s) that better correlate, for each case, with the proton transfer. The analysis might provide qualitative and quantitative description of the different system features responsible for the significant difference in the reported rates.

For the two systems, we performed different independent simulations using different initialization procedures available within the PyRETIS package \cite{riccardi2020pyretis} to reduce the correlation in the exploration of path space and to speed up sampling. The results obtained from each independent simulation have been then combined. Even if the analysis of individual simulations further demonstrated the efficiency of our method to reveal partial difference within realizations (e.g. different, less likely, reaction channels), we limit ourselves to discussion of the findings from the total combined data in the present manuscript.  

Generally, all trajectory segments, or trajectories for path sampling, can be considered in the present analysis approach. When using the latter simulation approach, a re-weight algorithm shall be adopted to consider all the generated paths. Due to the statistical weight of the different ensemble, and for simplicity, we opted to consider only the trajectories included in the  the outermost ensemble in path space (for the definition of an ensemble please consider the RETIS formalism \cite{vanErp2003a}).

Each considered trajectory, or trajectory segment, has to be labeled according to a reactive/unreactive definition. The frames from the selection windows have been transformed into the distance matrix representation (transnational and rotational invariant) and fed to the machine learning algorithm to generate the forthcoming analysis. We want to stress here that these results are generated without any arbitrary hypothesis or bias. The only hyper-parameters are the number of frames per trajectory, the location and size of the selection window. 

Decision trees, visualisations and random forest of decision trees to assert the first split error have been generated for the two systems. We here report the results obtained by the index-invariant distance matrix, which is the most general approach, even if more computationally demanding. It is worth noting that the index-variant distance matrix can be still advantageous for its simplicity and symmetry in certain applications. For example, in the presence of atoms that do now swap order during a transition. In the SI, the results for the index-variant distance matrix will also be presented.

The atom labelling system we use identifies each atom with a character and a digit. The character corresponds to the atom type, while the digit corresponds to the position of the sorted distance list per element with respect to a reference atom, with the indexing starting at 0.

The digit of the first entry in the atom-atom distance label refers to the sorted distance list with respect to the reference atom (C). The digit of the second entry refers to the sorted distance list with respect to the first atom of the atom-atom pair.
As an example, O2-H5 corresponds to the distance of the third closest oxygen (O2) from the C atom to the hydrogen atom which is 6$^{th}$ closest to the  oxygen in the pair (O2).  H0-O0 is the distance from the H closest to the C (H0) to the oxygen closest to H0.

\subsection{Formic Acid with 4 water molecules}

\begin{figure}
    \centering 
    \includegraphics[width=\textwidth]{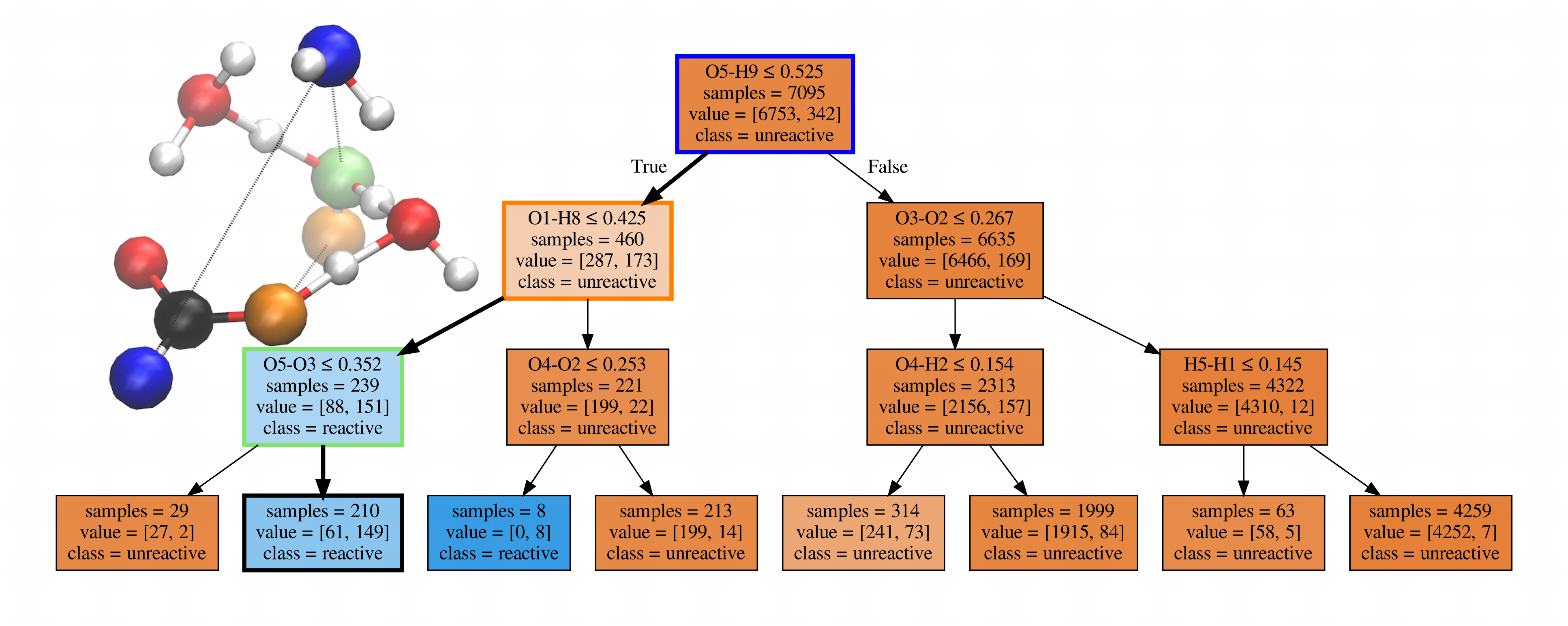} 
    \caption{Decision tree for the system with 4 water molecules around the formic acid molecule based on the index-invariant distance matrix. Three splits divide the input, each square reports 1) the question to split the data, 2) the number of samples going into the node, 3) the number of [unreactive, reactive] samples going into the node and 4) the majority class of the node. At each split, the True branch is on the left, False on the right. The color indicates the ratio between unreactive (brown) and reactive (blue) samples included in a node. Wider arrows have been used to link the decision three split with the atoms involved. In the top left corner, a 3D representation of the system is provided. The atoms highlighted in blue, yellow, and green, correspond to the atoms involved in the first, second, and third split of the decision three, respectively. In red, white and back are the oxygen, hydrogen and carbon atoms not indicated by the decision tree.} 
    \label{fig:4w_tree_ok}
\end{figure}

The decision tree generated for the four water molecules cluster around the FA is reported in figure \ref{fig:4w_tree_ok}. To simplify the visualization of the main splits that lead to the highest reactive trajectories of the decision tree, in the figure \ref{fig:4w_tree_ok}, a \textit{xyz} representation has been included. The atoms in blue, yellow, and green are involved in the first, second, and third split, respectively. 

The deprotonation reaction of FA appears to primarily require the distance between O5 and H9 being  smaller than 5.25~Å. This split implies that the distance between the furthest oxygen and the furthest hydrogen from such oxygen should be within a given threshold.  As H9 is the hydrogen of FA, that also implies that a certain orientation of the molecule, in respect to the water cluster, is also required. In these conditions, the probability for the path to be reactive is 38\%.

The next split, along the branch which leads to the highest reactive stance, is the distance between O1 and H8 being smaller than 4.25~Å. The distance between one of the FA oxygens and one of the furthest hydrogen should be sufficiently small. That implies that the oxygen of the FA should be located around the center of the cluster and that a sort of specific structure of the water cluster is required to satisfy this condition. With this condition, the probability for a path to be reactive  reaches 63\%  

Still along the branch towards the highest reactive stance, the distance between O5 and 03 being bigger than 3.52~Å represent the last split here considered. This corresponds to the relative position of two water molecules being two hydrogen bonds away. We interpret this requisite as the suitable distance to establish hydrogen bonding between the atoms. When the three requirements are met, the probability of for a path of being reactive is of 71\%

\begin{figure}
    \centering
    \includegraphics[width=0.92\textwidth]{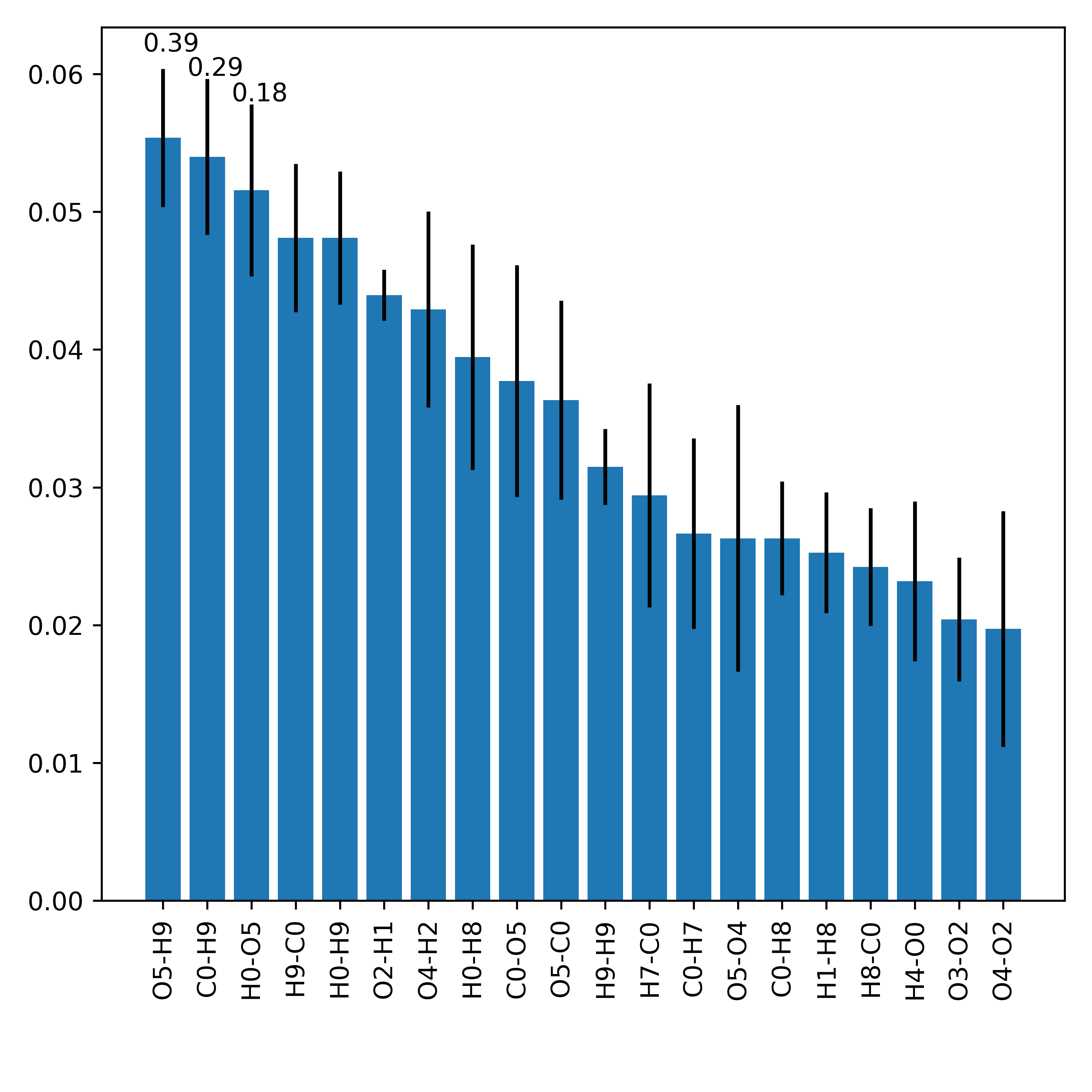}
    \caption{The importance approximations of the first split question from a random forest with depth of one for the four water case system. 
    The bars represent the feature importance of a Random Forest, with the error bar calculated with a block-error average based on the generated trajectories. The probability that a split is the true most important split is shown above the bars for the 3 most probable first splits.} 
    \label{fig:4w_errors_ok}
\end{figure}

By comparing the number of reactive paths versus the number of unreactive paths in the final splits of the decision tree, it can be concluded that the indicated reactive path  is clearly predominant. A similar conclusion can be reached by observing the first splits reported in figure~\ref{fig:4w_errors_ok}. The figure, which reports the results obtained from a random forest of decision trees of depth 1, indicates the relative probability for the first split to be the most important feature with a 39\% probability. The following distances reported by the random forest have a constantly decaying relevance. The first 5 main splits reported by figure~\ref{fig:4w_errors_ok} are correlated, and indicate that the water cluster has to be sufficiently compact and the FA has to be oriented such that its oxygen are in contact with the surrounding water molecules.

\subsection{Formic Acid with 6 water molecules}

\begin{figure}
    \centering 
    \includegraphics[width=\textwidth]{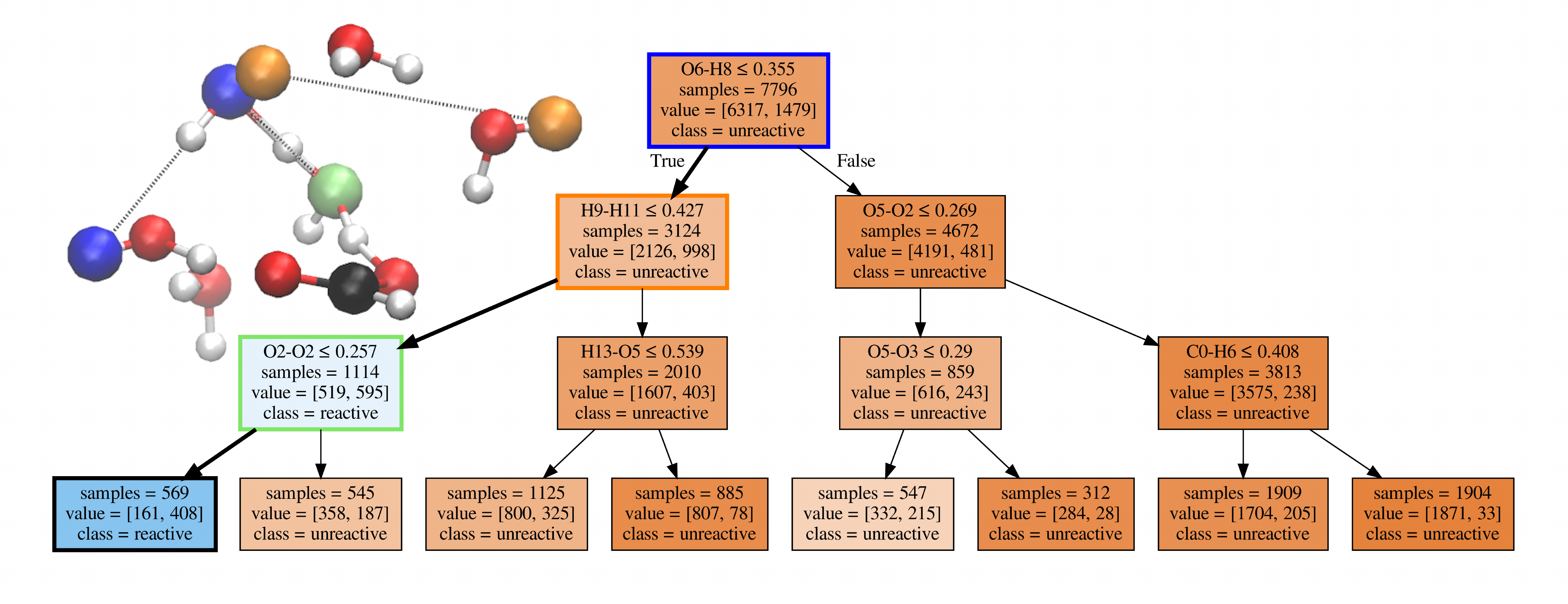}  
    \caption{Decision tree for the system with 6 water molecules around the formic acid molecule. Three splits divide the input, each square reports 1) the question to split the data, 2) the number of samples going into the node, 3) the number of [unreactive, reactive] samples going into the node and 4) the majority class of the node. At each split, the True branch is on the left, False on the right. The color indicates the ratio between unreactive (brown) and reactive (blue) samples included in a node. Wider arrows have been used to link the decision three split with the atoms involved. In the top left corner, a 3D representation of the system is provided. The atoms highlighted in blue, yellow, and green, correspond to the atoms involved in the first, second, and third split of the decision three, respectively.  In red, white and back are the oxygen, hydrogen and carbon atoms not indicated by the decision tree.} 
    \label{fig:6w_tree_ok}
\end{figure}

For the six water molecules water clusters around FA, in figure \ref{fig:6w_tree_ok}, we report the generated decision tree. Consistently with the four water molecule case, a visualization of the main splits of the decision tree that lead to the highest reactive trajectories is also included in figure \ref{fig:6w_tree_ok}. The atoms in blue, yellow, and green are involved in the first, second, and third slit, respectively.

The deprotonation reaction of FA in the six water molecule cluster requires a distance between O6 and H8 being  smaller than 3.55~Å. This split involves two water molecules in the proximity of FA, that needs to be within a certain distance. By inspecting the frame reported in figure \ref{fig:6w_tree_ok}, the requirement seems to indicate a water structure in the surrounding of the water molecule in proximity with the FA oxygen.  In such conditions the probability for the path to be reactive is 32\%.

The next split, along the branch that leads to the highest reactive stance, is the distance between H9 and H11 being smaller than 4.27~Å. The distance between these two atoms can also be interpreted as a combination of molecular orientation of the water molecules in the surrounding of the FA and water cluster size. The probability of a reactive path reaches 53\% when such conditions occur.

Still along the branch towards the highest reactive stance, the distance between O2 and 02 being bigger than 2.57~Å represent the last split here considered. This indicates that the closest water oxygen to FA (O2) should be close enough to a second oxygens to promote the formation of a hydrogen bond network. When the three requirements are met, the probability for a path of being reactive is of 72\%

\begin{figure}
    \centering
    \includegraphics[width=0.92\textwidth]{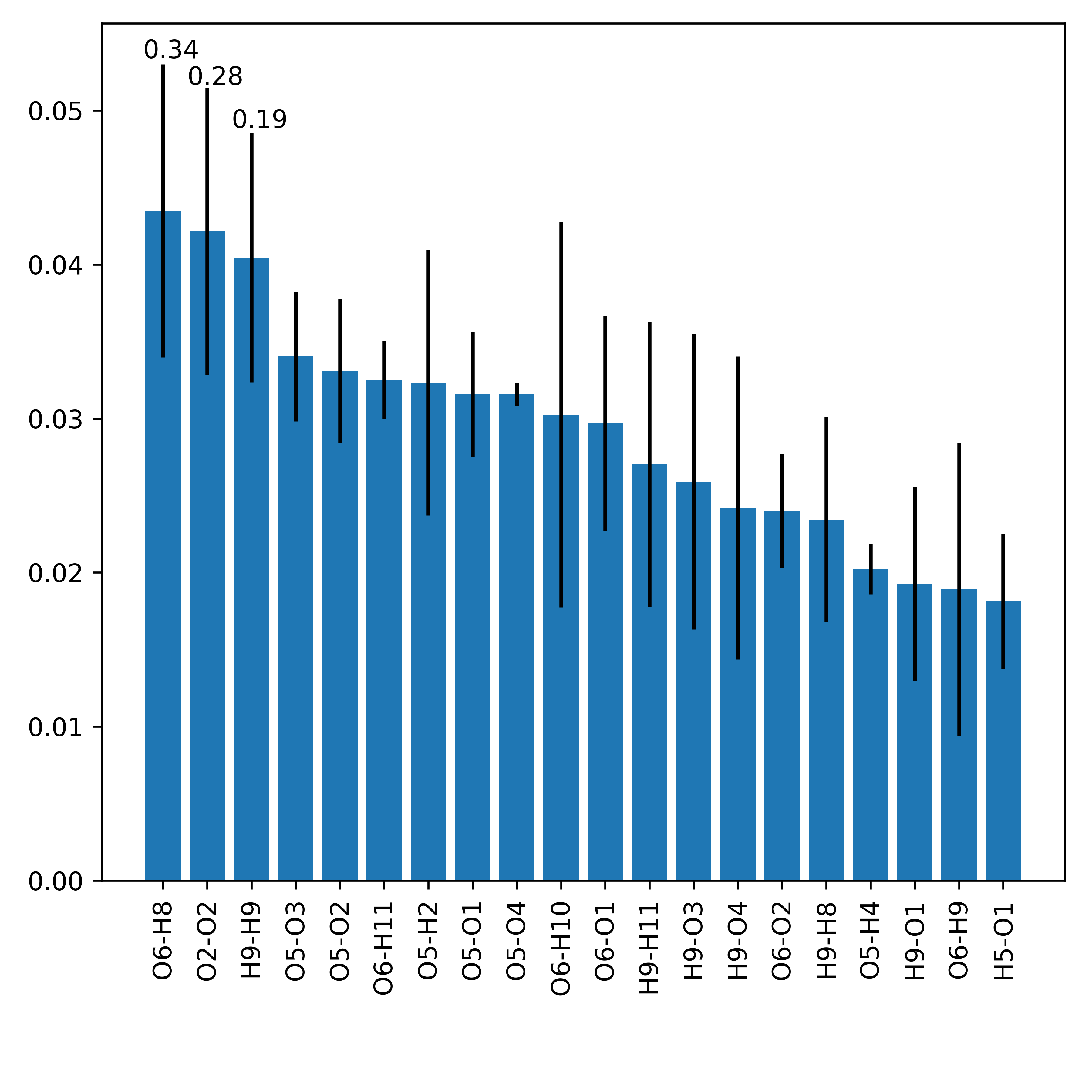}
    \caption{The importance approximations of the first split question from a random forest with depth of one for the 6 water case system. 
    The bars represent the feature importance of a Random Forest, with the error bar calculated with a block-error average based on the generated trajectories. The probability that a split is the true most important split is shown above the bars for the 3 most likely first splits.} 
    \label{fig:6w_errors_ok}
\end{figure}

By comparing the number of reactive paths versus the number of unreactive paths in the final splits of the decision tree, it can be concluded that the indicated reactive path is clearly favourable, but that other significant paths can also exist. This conclusion is also supported by the random forest of decision trees with a single split. Before proposing an interpretation, it is worth to remind here that the random forest reports unconditional entries, while the decision tree splits depending from the first split.  Figure~\ref{fig:6w_errors_ok} indicates the first split to be the most important feature is 34\%, while the second has a comparable relevance: O2-O2 (28\%). This confirms that, while a predominant pathway for the reaction has been sampled, different main pathways can co-exist.

As reported in figure \ref{fig:kload}, the dimension and the water cluster had a significant effect on the rate of the proton transfer reaction. From the comparison of the previously discussed figures \ref{fig:4w_tree_ok} and \ref{fig:6w_tree_ok}, it can be noted that the distance between an oxygen and a hydrogen in the furthermost regions in respect to the FA is the predominant characteristic for a trajectory to be reactive. Both clusters have to be sufficiently compact in order to promote the reaction. In the four water molecule case, the orientation of the FA in respect of the water cluster is the most important feature, while for the six water molecule case the water structure around the FA appears to be the predominant feature.

A second main difference between the four and six water cases is the possible pathways for the reaction to occur. The smaller system has a predominant reactive path, while for the six water molecule cluster, multiple paths appear to co-exist, contributing to the final reaction rate. Physically, if the system is sufficiently large, different configurations can lead to the proton transfer reaction.

\section{Conclusions}

A data driven method to systematically compute reaction pathways has been presented. The conventional \textit{xyz} data representation employed in molecular simulations is converted into an index invariant distance matrix representation, which is also translation and rotation invariant. Thereafter, an approach to limit the correlation between source data (molecular dynamic trajectories) has been also proposed, in consistency with rare event simulation framework. The data has then fed to a supervised classifier method, the decision tree.
To simplify the interpretation of the classifier, a back mapping procedure from the index invariant matrix has been adopted to emphasize the atoms involved with each split identified by the decision tree. Random forest of decision trees, in combination with block averaging, provided an error range of the first split of the decision tree.

We thus presented a data-driven approach to gain insight into a chemical reaction. The method has been designed such to be readily applicable  to other simulation strategies and transitions.  The strength of the present approach is to generate complex collective variables, and estimate the probability of their occurrence in a transition path. The descriptors to elucidate transition mechanisms  might be directly implemented in a prediction method.~\cite{van2016analyzing}

The method adopted an index invariant distance matrix providing a data driven insight into the reaction pathways. The data driven identification aims to identify interpretable pathways in a system composed of indistinguishable molecules. Applications to more inhomogeneous systems would be straightforward, especially if only a portion of the system atom is of interest. The latter case would combine a human intuition with a data driven approach, which would, possibly, provide a better insight into the reaction if, and only if, the introduced bias is correct. Our method can be further expanded by considering a higher number of descriptors alongside the distance matrix. Velocities, angles between molecules, coarse grain procedure or a mix of user defined functions,~\cite{ola2021} can be fed to the decision tree and follow up analysis.

To demonstrate the capabilities of the developed method, a mechanistic description of the proton transfer reaction has been provided. The reaction has been simulated via rare event simulation (replica exchange transition interface sampling \cite{vanErp2007a}) and its rate quantified for two water clusters, one composed of four and one of six water molecules surrounding a FA molecule. 

The reaction rate we measure is strongly influenced by the number of water molecules present. Mechanistically, the four and six water proton transfer reaction requires the water cluster to be sufficiently compact. The four water molecules system requires a certain orientation of the FA molecules and of the water molecules in its proximity. For the six water molecule case, a certain orientation of the outer water molecules appears to be more significant in describing the reaction path.
Furthermore, the four water cluster system indicated a predominant pathway for the reaction to occur, while in the six water molecules cluster, several pathways have been identified contributing to the reaction.

\begin{acknowledgement}
Part of the work has been performed under the Project HPC-EUROPA3 (INFRAIA-2016-1-730897), with the support of the EC Research Innovation Action under the H2020 Programme; 
in particular, the authors gratefully acknowledge the support of the Department of Chemistry at the University of Helsinki and the computer resources and technical support provided by CSC.  Christopher D. Daub acknowledges funding by the Academy of Finland (Grant Number 294752) and by the Jane and Aatos Erkko Foundation.

\end{acknowledgement}

\bibliography{chemistrees}

\providecommand{\latin}[1]{#1}
\makeatletter
\providecommand{\doi}
  {\begingroup\let\do\@makeother\dospecials
  \catcode`\{=1 \catcode`\}=2 \doi@aux}
\providecommand{\doi@aux}[1]{\endgroup\texttt{#1}}
\makeatother
\providecommand*\mcitethebibliography{\thebibliography}
\csname @ifundefined\endcsname{endmcitethebibliography}
  {\let\endmcitethebibliography\endthebibliography}{}
\begin{mcitethebibliography}{39}
\providecommand*\natexlab[1]{#1}
\providecommand*\mciteSetBstSublistMode[1]{}
\providecommand*\mciteSetBstMaxWidthForm[2]{}
\providecommand*\mciteBstWouldAddEndPuncttrue
  {\def\EndOfBibitem{\unskip.}}
\providecommand*\mciteBstWouldAddEndPunctfalse
  {\let\EndOfBibitem\relax}
\providecommand*\mciteSetBstMidEndSepPunct[3]{}
\providecommand*\mciteSetBstSublistLabelBeginEnd[3]{}
\providecommand*\EndOfBibitem{}
\mciteSetBstSublistMode{f}
\mciteSetBstMaxWidthForm{subitem}{(\alph{mcitesubitemcount})}
\mciteSetBstSublistLabelBeginEnd
  {\mcitemaxwidthsubitemform\space}
  {\relax}
  {\relax}

\bibitem[Murdachaew \latin{et~al.}(2016)Murdachaew, Nathanson, Gerber, and
  Halonen]{Murdachaew2016a}
Murdachaew,~G.; Nathanson,~G.~M.; Gerber,~R.~B.; Halonen,~L. Deprotonation of
  formic acid in collisions with a liquid water surface studied by molecular
  dynamics and metadynamics simulations. \emph{Phys.~Chem.~Chem.~Phys.}
  \textbf{2016}, \emph{18}, 29756--29770\relax
\mciteBstWouldAddEndPuncttrue
\mciteSetBstMidEndSepPunct{\mcitedefaultmidpunct}
{\mcitedefaultendpunct}{\mcitedefaultseppunct}\relax
\EndOfBibitem
\bibitem[Millet \latin{et~al.}(2015)Millet, Baasandorj, Farmer, Thornton,
  Baumann, Brophy, Chaliyakunnel, de~Gouw, Graus, Hu, Koss, Lee,
  Lopez-Hilfiker, Neuman, Paulot, Peischl, Pollack, Ryerson, Warneke, Williams,
  and Xu]{Millet2015a}
Millet,~D.~B. \latin{et~al.}  A large and ubiquitous source of atmospheric
  formic acid. \emph{Atmos.~Chem.~Phys.} \textbf{2015}, \emph{15},
  6283--6304\relax
\mciteBstWouldAddEndPuncttrue
\mciteSetBstMidEndSepPunct{\mcitedefaultmidpunct}
{\mcitedefaultendpunct}{\mcitedefaultseppunct}\relax
\EndOfBibitem
\bibitem[Chaliyakunnel \latin{et~al.}(2016)Chaliyakunnel, Millet, Wells,
  Cady-Pereira, and Shephard]{Chaliyakunnel2016a}
Chaliyakunnel,~S.; Millet,~D.~B.; Wells,~K.~C.; Cady-Pereira,~K.~E.;
  Shephard,~M.~W. A Large Underestimate of Formic Acid from Tropical Fires:
  Constraints from Space-Borne Measurements. \emph{Environ.~Sci.~Technol.}
  \textbf{2016}, \emph{50}, 5631--5640\relax
\mciteBstWouldAddEndPuncttrue
\mciteSetBstMidEndSepPunct{\mcitedefaultmidpunct}
{\mcitedefaultendpunct}{\mcitedefaultseppunct}\relax
\EndOfBibitem
\bibitem[Kangas \latin{et~al.}(2020)Kangas, H{\"a}nninen, and
  Halonen]{Kangas2020a}
Kangas,~P.; H{\"a}nninen,~V.; Halonen,~L. An Ab Initio Molecular Dynamics Study
  of the Hydrolysis Reaction of Sulfur Trioxide Catalyzed by a Formic Acid or
  Water Molecule. \emph{J.~Phys.~Chem.~A} \textbf{2020}, \relax
\mciteBstWouldAddEndPunctfalse
\mciteSetBstMidEndSepPunct{\mcitedefaultmidpunct}
{}{\mcitedefaultseppunct}\relax
\EndOfBibitem
\bibitem[Daub \latin{et~al.}(2020)Daub, Riccardi, H\"{a}nninen, and
  Halonen]{Daub2020a}
Daub,~C.~D.; Riccardi,~E.; H\"{a}nninen,~V.; Halonen,~L. Path sampling for
  atmospheric reactions: formic acid catalysed conversion of {SO}3 $+$ H2O to
  H2SO4. \emph{{PeerJ} Physical Chemistry} \textbf{2020}, \emph{2}, e7\relax
\mciteBstWouldAddEndPuncttrue
\mciteSetBstMidEndSepPunct{\mcitedefaultmidpunct}
{\mcitedefaultendpunct}{\mcitedefaultseppunct}\relax
\EndOfBibitem
\bibitem[Daub and Halonen(2019)Daub, and Halonen]{Daub2019b}
Daub,~C.~D.; Halonen,~L. Ab Initio Molecular Dynamics Simulations of the
  Influence of Lithium Bromide Salt on the Deprotonation of Formic Acid in
  Aqueous Solution. \emph{J.~Phys.~Chem.~B} \textbf{2019}, \emph{123},
  6823--6829\relax
\mciteBstWouldAddEndPuncttrue
\mciteSetBstMidEndSepPunct{\mcitedefaultmidpunct}
{\mcitedefaultendpunct}{\mcitedefaultseppunct}\relax
\EndOfBibitem
\bibitem[van Erp(2007)]{vanErp2007a}
van Erp,~T.~S. Reaction Rate Calculation by Parallel Path Swapping.
  \emph{Phys.~Rev.~Lett.} \textbf{2007}, \emph{98}, 268301\relax
\mciteBstWouldAddEndPuncttrue
\mciteSetBstMidEndSepPunct{\mcitedefaultmidpunct}
{\mcitedefaultendpunct}{\mcitedefaultseppunct}\relax
\EndOfBibitem
\bibitem[Riccardi \latin{et~al.}(2017)Riccardi, Dahlen, and van
  Erp]{Riccardi2017a}
Riccardi,~E.; Dahlen,~O.; van Erp,~T.~S. Fast Decorrelating {Monte Carlo} Moves
  for Efficient Path Sampling. \emph{J.~Phys.~Chem.~Lett.} \textbf{2017},
  \emph{8}, 4456--4460\relax
\mciteBstWouldAddEndPuncttrue
\mciteSetBstMidEndSepPunct{\mcitedefaultmidpunct}
{\mcitedefaultendpunct}{\mcitedefaultseppunct}\relax
\EndOfBibitem
\bibitem[Riccardi \latin{et~al.}(2019)Riccardi, Van~Mastbergen, Navarre, and
  Vreede]{riccardi2019predicting}
Riccardi,~E.; Van~Mastbergen,~E.~C.; Navarre,~W.~W.; Vreede,~J. Predicting the
  mechanism and rate of H-NS binding to AT-rich DNA. \emph{PLoS computational
  biology} \textbf{2019}, \emph{15}, e1006845\relax
\mciteBstWouldAddEndPuncttrue
\mciteSetBstMidEndSepPunct{\mcitedefaultmidpunct}
{\mcitedefaultendpunct}{\mcitedefaultseppunct}\relax
\EndOfBibitem
\bibitem[Riccardi \latin{et~al.}(2020)Riccardi, Kr\"amer, van Erp, and
  Ghysels]{riccardi2020permeation}
Riccardi,~E.; Kr\"amer,~A.; van Erp,~T.~S.; Ghysels,~A. Permeation Rates of
  Oxygen through a Lipid Bilayer Using Replica Exchange Transition Interface
  Sampling. \emph{The Journal of Physical Chemistry B} \textbf{2020}, \relax
\mciteBstWouldAddEndPunctfalse
\mciteSetBstMidEndSepPunct{\mcitedefaultmidpunct}
{}{\mcitedefaultseppunct}\relax
\EndOfBibitem
\bibitem[Moqadam \latin{et~al.}(2017)Moqadam, Riccardi, Trinh, Lervik, and van
  Erp]{moqadam2017rare}
Moqadam,~M.; Riccardi,~E.; Trinh,~T.~T.; Lervik,~A.; van Erp,~T.~S. Rare event
  simulations reveal subtle key steps in aqueous silicate condensation.
  \emph{Physical Chemistry Chemical Physics} \textbf{2017}, \emph{19},
  13361--13371\relax
\mciteBstWouldAddEndPuncttrue
\mciteSetBstMidEndSepPunct{\mcitedefaultmidpunct}
{\mcitedefaultendpunct}{\mcitedefaultseppunct}\relax
\EndOfBibitem
\bibitem[Moqadam \latin{et~al.}(2018)Moqadam, Lervik, Riccardi, Venkatraman,
  Alsberg, and van Erp]{Moqadam2018a}
Moqadam,~M.; Lervik,~A.; Riccardi,~E.; Venkatraman,~V.; Alsberg,~B.~K.; van
  Erp,~T.~S. Local initiation conditions for water autoionization.
  \emph{Proc.~Nat.~Acad.~Sci.} \textbf{2018}, \emph{115}, E4569--E4576\relax
\mciteBstWouldAddEndPuncttrue
\mciteSetBstMidEndSepPunct{\mcitedefaultmidpunct}
{\mcitedefaultendpunct}{\mcitedefaultseppunct}\relax
\EndOfBibitem
\bibitem[van Erp \latin{et~al.}(2016)van Erp, Moqadam, Riccardi, and
  Lervik]{van2016analyzing}
van Erp,~T.~S.; Moqadam,~M.; Riccardi,~E.; Lervik,~A. Analyzing complex
  reaction mechanisms using path sampling. \emph{Journal of Chemical Theory and
  Computation} \textbf{2016}, \emph{12}, 5398--5410\relax
\mciteBstWouldAddEndPuncttrue
\mciteSetBstMidEndSepPunct{\mcitedefaultmidpunct}
{\mcitedefaultendpunct}{\mcitedefaultseppunct}\relax
\EndOfBibitem
\bibitem[Hooft \latin{et~al.}(2021)Hooft, de~Alba~Ort{\'{\i}}z, and
  Ensing]{Hooft2021}
Hooft,~F.; de~Alba~Ort{\'{\i}}z,~A.~P.; Ensing,~B. Discovering Collective
  Variables of Molecular Transitions via Genetic Algorithms and Neural
  Networks. \emph{Journal of Chemical Theory and Computation} \textbf{2021},
  \relax
\mciteBstWouldAddEndPunctfalse
\mciteSetBstMidEndSepPunct{\mcitedefaultmidpunct}
{}{\mcitedefaultseppunct}\relax
\EndOfBibitem
\bibitem[Chen and Ferguson(2018)Chen, and Ferguson]{Chen2018}
Chen,~W.; Ferguson,~A.~L. Molecular enhanced sampling with autoencoders:
  On-the-fly collective variable discovery and accelerated free energy
  landscape exploration. \emph{Journal of Computational Chemistry}
  \textbf{2018}, \emph{39}, 2079--2102\relax
\mciteBstWouldAddEndPuncttrue
\mciteSetBstMidEndSepPunct{\mcitedefaultmidpunct}
{\mcitedefaultendpunct}{\mcitedefaultseppunct}\relax
\EndOfBibitem
\bibitem[Sch\"{o}berl \latin{et~al.}(2019)Sch\"{o}berl, Zabaras, and
  Koutsourelakis]{Schberl2019}
Sch\"{o}berl,~M.; Zabaras,~N.; Koutsourelakis,~P.-S. Predictive collective
  variable discovery with deep Bayesian models. \emph{The Journal of Chemical
  Physics} \textbf{2019}, \emph{150}, 024109\relax
\mciteBstWouldAddEndPuncttrue
\mciteSetBstMidEndSepPunct{\mcitedefaultmidpunct}
{\mcitedefaultendpunct}{\mcitedefaultseppunct}\relax
\EndOfBibitem
\bibitem[Ribeiro \latin{et~al.}(2018)Ribeiro, Bravo, Wang, and
  Tiwary]{Ribeiro2018}
Ribeiro,~J. M.~L.; Bravo,~P.; Wang,~Y.; Tiwary,~P. Reweighted autoencoded
  variational Bayes for enhanced sampling ({RAVE}). \emph{The Journal of
  Chemical Physics} \textbf{2018}, \emph{149}, 072301\relax
\mciteBstWouldAddEndPuncttrue
\mciteSetBstMidEndSepPunct{\mcitedefaultmidpunct}
{\mcitedefaultendpunct}{\mcitedefaultseppunct}\relax
\EndOfBibitem
\bibitem[Jung \latin{et~al.}(2019)Jung, Covino, and Hummer]{jung2019}
Jung,~H.; Covino,~R.; Hummer,~G. Artificial intelligence assists discovery of
  reaction coordinates and mechanisms from molecular dynamics simulations.
  \emph{arXiv preprint arXiv:1901.04595} \textbf{2019}, \relax
\mciteBstWouldAddEndPunctfalse
\mciteSetBstMidEndSepPunct{\mcitedefaultmidpunct}
{}{\mcitedefaultseppunct}\relax
\EndOfBibitem
\bibitem[Swain and Hauska(1977)Swain, and Hauska]{swain1977decision}
Swain,~P.~H.; Hauska,~H. The decision tree classifier: Design and potential.
  \emph{IEEE Transactions on Geoscience Electronics} \textbf{1977}, \emph{15},
  142--147\relax
\mciteBstWouldAddEndPuncttrue
\mciteSetBstMidEndSepPunct{\mcitedefaultmidpunct}
{\mcitedefaultendpunct}{\mcitedefaultseppunct}\relax
\EndOfBibitem
\bibitem[Leopold(2011)]{Leopold2011a}
Leopold,~K.~R. Hydrated Acid Clusters. \emph{Annu.~Rev.~Phys.~Chem.}
  \textbf{2011}, \emph{62}, 327--349\relax
\mciteBstWouldAddEndPuncttrue
\mciteSetBstMidEndSepPunct{\mcitedefaultmidpunct}
{\mcitedefaultendpunct}{\mcitedefaultseppunct}\relax
\EndOfBibitem
\bibitem[Forbert \latin{et~al.}(2011)Forbert, Masia, Kaczmarek-Kedziera, Nair,
  and Marx]{Forbert2011a}
Forbert,~H.; Masia,~M.; Kaczmarek-Kedziera,~A.; Nair,~N.~N.; Marx,~D.
  Aggregation-Induced Chemical Reactions: Acid Dissociation in Growing Water
  Clusters. \emph{J.~Am.~Chem.~Soc.} \textbf{2011}, \emph{133},
  4062--4072\relax
\mciteBstWouldAddEndPuncttrue
\mciteSetBstMidEndSepPunct{\mcitedefaultmidpunct}
{\mcitedefaultendpunct}{\mcitedefaultseppunct}\relax
\EndOfBibitem
\bibitem[Chung and Kim(2017)Chung, and Kim]{Chung2017a}
Chung,~Y.~K.; Kim,~S.~K. Dissociation of sulfur oxoacids by two water molecules
  studied using ab initio and density functional theory calculations.
  \emph{Int.~J.~Quantum~Chem.} \textbf{2017}, \emph{117}, e25419\relax
\mciteBstWouldAddEndPuncttrue
\mciteSetBstMidEndSepPunct{\mcitedefaultmidpunct}
{\mcitedefaultendpunct}{\mcitedefaultseppunct}\relax
\EndOfBibitem
\bibitem[Lengyel \latin{et~al.}(2017)Lengyel, Pysanenko, and
  F{\'a}rn{\'i}k]{Lengyel2017a}
Lengyel,~J.; Pysanenko,~A.; F{\'a}rn{\'i}k,~M. Electron-induced chemistry in
  microhydrated sulfuric acid clusters. \emph{Atmos.~Chem.~Phys.}
  \textbf{2017}, \emph{17}, 14171--14180\relax
\mciteBstWouldAddEndPuncttrue
\mciteSetBstMidEndSepPunct{\mcitedefaultmidpunct}
{\mcitedefaultendpunct}{\mcitedefaultseppunct}\relax
\EndOfBibitem
\bibitem[Gutberlet \latin{et~al.}(2009)Gutberlet, Schwaab, Birer, Masia,
  Kaczmarek, Forbert, Havenith, and Marx]{Gutberlet2009a}
Gutberlet,~A.; Schwaab,~G.; Birer,~{\" O}.; Masia,~M.; Kaczmarek,~A.;
  Forbert,~H.; Havenith,~M.; Marx,~D. Aggregation-Induced Dissociation of
  HCl(H$_2$O)$_4$ Below 1 K: The Smallest Droplet of Acid. \emph{Science}
  \textbf{2009}, \emph{324}, 1545\relax
\mciteBstWouldAddEndPuncttrue
\mciteSetBstMidEndSepPunct{\mcitedefaultmidpunct}
{\mcitedefaultendpunct}{\mcitedefaultseppunct}\relax
\EndOfBibitem
\bibitem[Maity(2013)]{Maity2013a}
Maity,~D.~K. How Much Water Is Needed To Ionize Formic Acid?
  \emph{J.~Phys.~Chem.~A} \textbf{2013}, \emph{117}, 8660--8670\relax
\mciteBstWouldAddEndPuncttrue
\mciteSetBstMidEndSepPunct{\mcitedefaultmidpunct}
{\mcitedefaultendpunct}{\mcitedefaultseppunct}\relax
\EndOfBibitem
\bibitem[Elena \latin{et~al.}(2013)Elena, Meloni, and Ciccotti]{Elena2013a}
Elena,~A.~M.; Meloni,~S.; Ciccotti,~G. Equilibrium and Rate Constants, and
  Reaction Mechanism of the HF Dissociation in the {HF(H$_2$O)$_7$} Cluster by
  ab Initio Rare Event Simulations. \emph{J.~Phys.~Chem.~A} \textbf{2013},
  \emph{117}, 13039--13050\relax
\mciteBstWouldAddEndPuncttrue
\mciteSetBstMidEndSepPunct{\mcitedefaultmidpunct}
{\mcitedefaultendpunct}{\mcitedefaultseppunct}\relax
\EndOfBibitem
\bibitem[VandeVondele \latin{et~al.}(2005)VandeVondele, Krack, Mohamed,
  Parrinello, Chassaing, and Hutter]{VanDeVondele2005a}
VandeVondele,~J.; Krack,~M.; Mohamed,~F.; Parrinello,~M.; Chassaing,~T.;
  Hutter,~J. {QUICKSTEP}: Fast and accurate density functional calculations
  using a mixed Gaussian and plane waves approach. \emph{Comput.~Phys.~Commun.}
  \textbf{2005}, \emph{167}, 103--128\relax
\mciteBstWouldAddEndPuncttrue
\mciteSetBstMidEndSepPunct{\mcitedefaultmidpunct}
{\mcitedefaultendpunct}{\mcitedefaultseppunct}\relax
\EndOfBibitem
\bibitem[Grimme \latin{et~al.}(2010)Grimme, Antony, Ehrlich, and
  Krieg]{grimme2010consistent}
Grimme,~S.; Antony,~J.; Ehrlich,~S.; Krieg,~H. A consistent and accurate ab
  initio parametrization of density functional dispersion correction (DFT-D)
  for the 94 elements H-Pu. \emph{The Journal of chemical physics}
  \textbf{2010}, \emph{132}, 154104\relax
\mciteBstWouldAddEndPuncttrue
\mciteSetBstMidEndSepPunct{\mcitedefaultmidpunct}
{\mcitedefaultendpunct}{\mcitedefaultseppunct}\relax
\EndOfBibitem
\bibitem[Lervik \latin{et~al.}(2017)Lervik, Riccardi, and van
  Erp]{lervik2017pyretis}
Lervik,~A.; Riccardi,~E.; van Erp,~T.~S. {PyRETIS:} A well-done, medium-sized
  python library for rare events. \emph{Journal of Computational Chemistry}
  \textbf{2017}, \emph{38}, 2439--2451\relax
\mciteBstWouldAddEndPuncttrue
\mciteSetBstMidEndSepPunct{\mcitedefaultmidpunct}
{\mcitedefaultendpunct}{\mcitedefaultseppunct}\relax
\EndOfBibitem
\bibitem[Riccardi \latin{et~al.}(2020)Riccardi, Lervik, Roet, Aar{\o}en, and
  van Erp]{riccardi2020pyretis}
Riccardi,~E.; Lervik,~A.; Roet,~S.; Aar{\o}en,~O.; van Erp,~T.~S. PyRETIS 2: an
  improbability drive for rare events. \emph{Journal of computational
  chemistry} \textbf{2020}, \emph{41}, 370--377\relax
\mciteBstWouldAddEndPuncttrue
\mciteSetBstMidEndSepPunct{\mcitedefaultmidpunct}
{\mcitedefaultendpunct}{\mcitedefaultseppunct}\relax
\EndOfBibitem
\bibitem[van Erp \latin{et~al.}(2003)van Erp, Moroni, and Bolhuis]{vanErp2003a}
van Erp,~T.~S.; Moroni,~D.; Bolhuis,~P.~G. A novel path sampling method for the
  calculation of rate constants. \emph{J.~Chem.~Phys.} \textbf{2003},
  \emph{118}, 7762--7774\relax
\mciteBstWouldAddEndPuncttrue
\mciteSetBstMidEndSepPunct{\mcitedefaultmidpunct}
{\mcitedefaultendpunct}{\mcitedefaultseppunct}\relax
\EndOfBibitem
\bibitem[Baker and Hehre(1991)Baker, and Hehre]{baker1991geometry}
Baker,~J.; Hehre,~W.~J. Geometry optimization in cartesian coordinates: The end
  of the Z-matrix? \emph{Journal of computational chemistry} \textbf{1991},
  \emph{12}, 606--610\relax
\mciteBstWouldAddEndPuncttrue
\mciteSetBstMidEndSepPunct{\mcitedefaultmidpunct}
{\mcitedefaultendpunct}{\mcitedefaultseppunct}\relax
\EndOfBibitem
\bibitem[Breiman \latin{et~al.}(2017)Breiman, Friedman, Olshen, and
  Stone]{Breiman2017}
Breiman,~L.; Friedman,~J.~H.; Olshen,~R.~A.; Stone,~C.~J. \emph{Classification
  And Regression Trees}; Routledge, 2017\relax
\mciteBstWouldAddEndPuncttrue
\mciteSetBstMidEndSepPunct{\mcitedefaultmidpunct}
{\mcitedefaultendpunct}{\mcitedefaultseppunct}\relax
\EndOfBibitem
\bibitem[Cabriolu \latin{et~al.}(2017)Cabriolu, Refsnes, Bolhuis, and van
  Erp]{Cabriolu2017a}
Cabriolu,~R.; Refsnes,~K.~M.~S.; Bolhuis,~P.~G.; van Erp,~T.~S. Foundations and
  latest advances in replica exchange transition interface sampling.
  \emph{J.~Chem.~Phys.} \textbf{2017}, \emph{147}, 152722\relax
\mciteBstWouldAddEndPuncttrue
\mciteSetBstMidEndSepPunct{\mcitedefaultmidpunct}
{\mcitedefaultendpunct}{\mcitedefaultseppunct}\relax
\EndOfBibitem
\bibitem[Pedregosa \latin{et~al.}(2011)Pedregosa, Varoquaux, Gramfort, Michel,
  Thirion, Grisel, Blondel, Prettenhofer, Weiss, Dubourg, Vanderplas, Passos,
  Cournapeau, Brucher, Perrot, and Duchesnay]{scikit-learn}
Pedregosa,~F. \latin{et~al.}  Scikit-learn: Machine Learning in {P}ython.
  \emph{Journal of Machine Learning Research} \textbf{2011}, \emph{12},
  2825--2830\relax
\mciteBstWouldAddEndPuncttrue
\mciteSetBstMidEndSepPunct{\mcitedefaultmidpunct}
{\mcitedefaultendpunct}{\mcitedefaultseppunct}\relax
\EndOfBibitem
\bibitem[Young and Householder(1938)Young, and Householder]{Young1938}
Young,~G.; Householder,~A.~S. Discussion of a set of points in terms of their
  mutual distances. \emph{Psychometrika} \textbf{1938}, \emph{3}, 19--22\relax
\mciteBstWouldAddEndPuncttrue
\mciteSetBstMidEndSepPunct{\mcitedefaultmidpunct}
{\mcitedefaultendpunct}{\mcitedefaultseppunct}\relax
\EndOfBibitem
\bibitem[Humphrey \latin{et~al.}(1996)Humphrey, Dalke, and
  Schulten]{Humphrey1996a}
Humphrey,~W.; Dalke,~A.; Schulten,~K. {VMD} -- {V}isual {M}olecular {D}ynamics.
  \emph{J. Molec. Graphics} \textbf{1996}, \emph{14}, 33--38\relax
\mciteBstWouldAddEndPuncttrue
\mciteSetBstMidEndSepPunct{\mcitedefaultmidpunct}
{\mcitedefaultendpunct}{\mcitedefaultseppunct}\relax
\EndOfBibitem
\bibitem[Aar{\o}en \latin{et~al.}(2021)Aar{\o}en, Ki{\ae}r, and
  Riccardi]{ola2021}
Aar{\o}en,~O.; Ki{\ae}r,~H.; Riccardi,~E. {PyVisA: Visualization and Analysis
  of path sampling trajectories}. \emph{Journal of Computational Chemistry}
  \textbf{2021}, \emph{42}, 435--446\relax
\mciteBstWouldAddEndPuncttrue
\mciteSetBstMidEndSepPunct{\mcitedefaultmidpunct}
{\mcitedefaultendpunct}{\mcitedefaultseppunct}\relax
\EndOfBibitem
\end{mcitethebibliography}

\end{document}


\subsection{Making the data atom-index invariant}

In a condensed system, the atoms might swap order during a transition. An index invariant data representation is, therefore, not clearly advantageous since it requires extra processing and it cannot use a symmetric representation. Yet, as it constitutes the most general case, it has been considered in addition to the index-variant representation we use in the main text.

\begin{figure}
    \begin{flushleft}
    \begin{tabular}{l|rrrrr|}
        & C0 & H0 & O0 & O1 & H1\\
        \hline
        C0 & 0.0000 & 0.1109 & 0.1250 & 0.1321 & 0.2010 \\
        H0 & 0.1109 & 0.0000 & 0.2047 & 0.2013 & 0.2954 \\
        O0 & 0.1250 & 0.2047 & 0.0000 & 0.2311 & 0.2524 \\
        O1 & 0.1321 & 0.2013 & 0.2311 & 0.0000 & 0.1073 \\
        H1 & 0.2010 & 0.2954 & 0.2524 & 0.1073 & 0.0000 \\
        \hline
    \end{tabular}
    \end{flushleft}

    \begin{tabular}{l|rrrrr|}
        & C0 & H0 & O0 & O1 & H1\\
        \hline
        C0 & 0.0000 & 0.2010 & 0.1250 & 0.1321 & 0.1109 \\
        H0 & 0.2010 & 0.0000 & 0.2524 & 0.1073 & 0.2954 \\
        O0 & 0.1250 & 0.2524 & 0.0000 & 0.2311 & 0.2047 \\
        O1 & 0.1321 & 0.1073 & 0.2311 & 0.0000 & 0.2013 \\
        H1 & 0.1109 & 0.2954 & 0.2047 & 0.2013 & 0.0000 \\
        \hline
    \end{tabular}
    \hfill
    \includegraphics[width=0.3\linewidth, clip, trim=55cm 15cm 55cm 14cm]{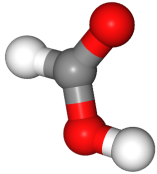}

    \caption{Two distance matrices (left) for an equivalent structure of formic acid (right). The difference between the two matrices is that H0 and H1 swapped places, or indices. This leads to a structure that has identical physics, but not an identical data representation. The distance matrix is therefore not a good representation to train our algorithms on during simulations where atom indices might change over time.}
    \label{fig:distissue}
\end{figure}

\begin{figure}
    \centering
    \begin{tabular}{l|rrrrr|}
        & C0 & H0 & H1 & O0 & O1 \\
        \hline
        C0 & 0.0000 & 0.1109 & 0.2010 & 0.1250 & 0.1321 \\
        H0 & 0.1109 & 0.0000 & 0.2954 & 0.2013 & 0.2047 \\
        H1 & 0.2010 & 0.0000 & 0.2954 & 0.1073 & 0.2524 \\
        O0 & 0.1250 & 0.2047 & 0.2524 & 0.0000 & 0.2311  \\
        O1 & 0.1321 & 0.1073 & 0.2013 & 0.0000 & 0.2311 \\
        \hline
    \end{tabular}
    \hfill
    \includegraphics[width=0.3\linewidth, clip, trim=55cm 15cm 55cm 14cm]{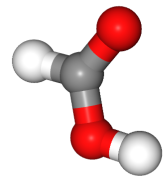}
    \caption{The index invariant distance matrix. It is created by first choosing an anchor point, in the present case, C0. Then the rows are grouped per element and sorted based on the distance from the anchor atom for each element. This representation is translationally, rotationally and atom-index invariant, making it a suitable general data representation to train ML algorithms on, including systems where atom indices might change.}
    \label{fig:distsolution}
\end{figure}

As shown in figure~\ref{fig:distissue}, by choosing a "anchor" atom, which may be different for each frame, the data representation can become invariant with respect to translation, rotation, and changes in the atomic indices. For the FA in the water system the carbon atom is the trivial identifiable anchor. The rest of the atoms are sorted based on the atom type and the distance from the anchor, as illustrated in figure~\ref{fig:distsolution}. The resulting data representation is atom-index invariant. Due to the statistical fluctuations in the atom positions, this procedure requires more data to achieve convergence of a ML algorithm, and imposes further requirements on the interpretation of the resulting random tree.

In our simulation, atoms of different indices do not swap places during transitions (They do, eventually, in the stable states). Therefore, we also reproduced our analysis with the relatively simple distance matrix.

Figure \ref{fig:mapping} reports a simplified algorithm to generate the distance matrix, while Figure \ref{fig:backmapping} presents the algorithm to generate the index invariant distance matrix.

\begin{algorithm}
    \caption{Element grouping}
    \begin{algorithmic}[1]
        \State mat = distance\_matrix
        \State out = output\_matrix
        \State indices = List()
        \For{element in Elements} \algorithmiccomment{group indices per element}
            \For{atom in atoms}
                \If{atom.element$=$element}
                \State{indices.append(atom.index)}
                \EndIf
            \EndFor
        \EndFor
        \State row\_out, col\_out$ = 0,0$
        \For{row\_i in indices} \algorithmiccomment{Group elements together}
            \For{col\_i in indices}
               \State out[row\_out][col\_out] = mat[row\_i][col\_i]
               \State col\_out $+= 1$
            \EndFor
            \State row\_out $+= 1$
        \EndFor
    \end{algorithmic}
    \caption{Algorithm to group the symmetric distance matrix based on elements. Inputs: 'mat'=a symmetric distance matrix, 'Elements'=a list of all elements in the system. Outputs: 'out'=a symmetric distance matrix where all atoms of the same element are grouped together.}
    \label{fig:mapping}
\end{algorithm}

\begin{algorithm}
    \caption{Sorting}
    \begin{algorithmic}[1]
        \State mat = grouped\_distance\_matrix \algorithmiccomment{Assume mat is grouped per element}
        \State out = output\_matrix
        \State elem\_length = List(count(atoms of element $e$) for $e \in $ Elements)
        \State out\_order = List()
        \State anchor\_idx = 0 \algorithmiccomment{Set anchor\_atom to row 0}
        \State anchor\_row = mat[anchor\_idx]
        \State i = 0
        \For{j in elem\_length} \algorithmiccomment{Figure out the output row-order.}
            \State out\_order.append(argsort(anchor\_row[i:i+j]))
            \State i $+=$ j
        \EndFor
        \State row\_out = 0
        \For{row\_idx in out\_order}
            \State row = mat[row\_idx]
            \State i = 0
            \For{j in elem\_length}
                \State out[row\_out][i:i+j] = sort(row[i:i+j])
                \State i $+=$ j
            \EndFor
            \State row\_out $+=$ 1
        \EndFor
    \end{algorithmic}
    \caption{Algorithm to make the element grouped distance matrix atom-index invariant.
    Inputs: 'mat'=a symmetric distance matrix that has been grouped per element, 'Elements'=a list of all elements in the system, 'anchor\_idx'=the row index which is the basis for the order of the rows in the output matrix. Outputs: 'out'=a sorted index-invariant distance matrix.}
    \label{fig:backmapping}
\end{algorithm}

\subsection{BackMapping the symmetric distance matrix to \textit{xyz}}

We adopted the procedure firstly suggested by \citet{Young1938}. If our distance matrix for a single frame is $D_{ij}$, we can construct the following mapping $M_{ij} = \frac{D_{1j}^{2}+D_{i1}^2 - D_{ij}^{2}}{2}$, where $D_{1j}$ is the $j$-th element of the first row of the distance matrix, and $D_{i1}$ is the $i$-th element of the first column.  

The eigenvalue decomposition on $M$ $M = USU^T$ allows the calculation of the matrix $X = U\sqrt{S}$. Only $N$ of the eigenvalues ($S$) are non-zero for a system that can be embedded in $N$ dimensional space, and distances are generated from a 3-dimensional space. Thus, the first 3 columns of $X$ corresponds to the \textit{x, y,} and \textit{z} coordinates for each row or column in the original distance matrix, up to a translation or rotation.

\section{Analysis based on the index variant distance-matrix}

As the result obtained by the index invariant data matrix has been included in the main paper, only the results obtained from the index variant case have been included here.

\subsection{Formic Acid with 4 water molecules}
 
 The frames identified by the selection windows have been transformed into the distance matrix representation (translational and rotational invariant) and fed to the machine learning algorithm to generate the forthcoming analysis. The only hyper-parameter is the location and size of the selection window.

\begin{figure}
    \centering
    \textbf{A}
    \includegraphics[width=\textwidth]{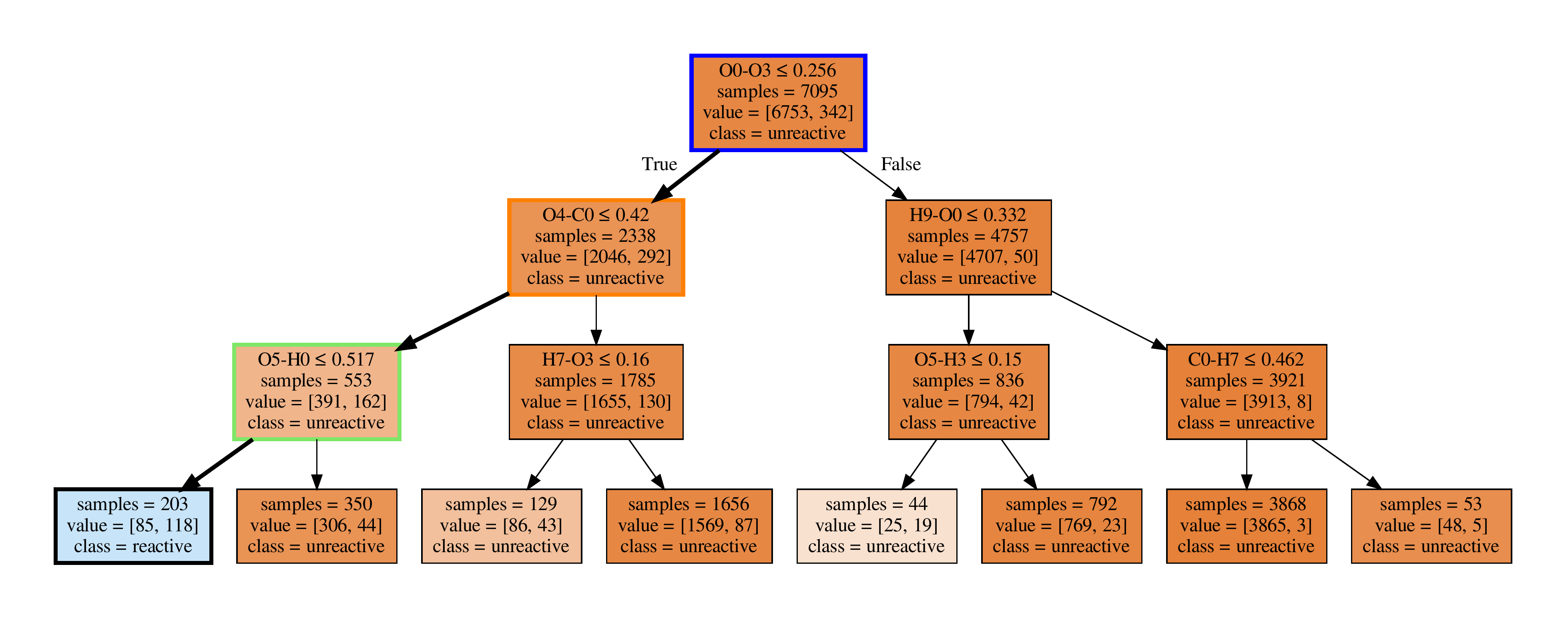}
    \\
    \textbf{B}
    \includegraphics[width=0.4\textwidth, clip, trim=3cm 4cm 4cm 5cm]{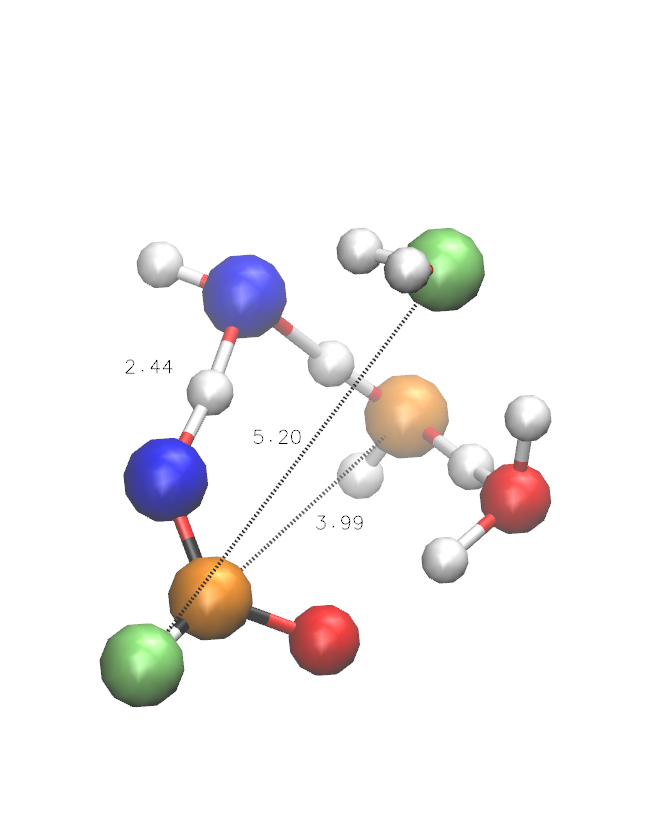}
    \caption{Decision tree for the system with 4 water molecules around the formic acid molecule based on the index-invariant distance matrix. Three splits divide the input, each square reports 1) the question to split the data, 2) the number of samples going into the node, 3) the number of [unreactive, reactive] samples going into the node and 4) the majority class of the node. At each split, the True branch is on the left, False on the right. The color indicates the ratio between unreactive (brown) and reactive (blue) samples included in a node. Wider arrows have been used to link the decision three split with the atoms involved. In panel B, a 3D representation of the system is provided. The atoms highlighted in blue, yellow, and green correspond to the atoms involved in the first, second, and third split of the decision three, respectively. In red are the oxygen and in white the hydrogen atoms not indicated by the decision tree reported in the panel B.}
    \label{fig:4w_tree_ok}
\end{figure}

\begin{figure}
    \centering
    \includegraphics[width=0.92\textwidth]{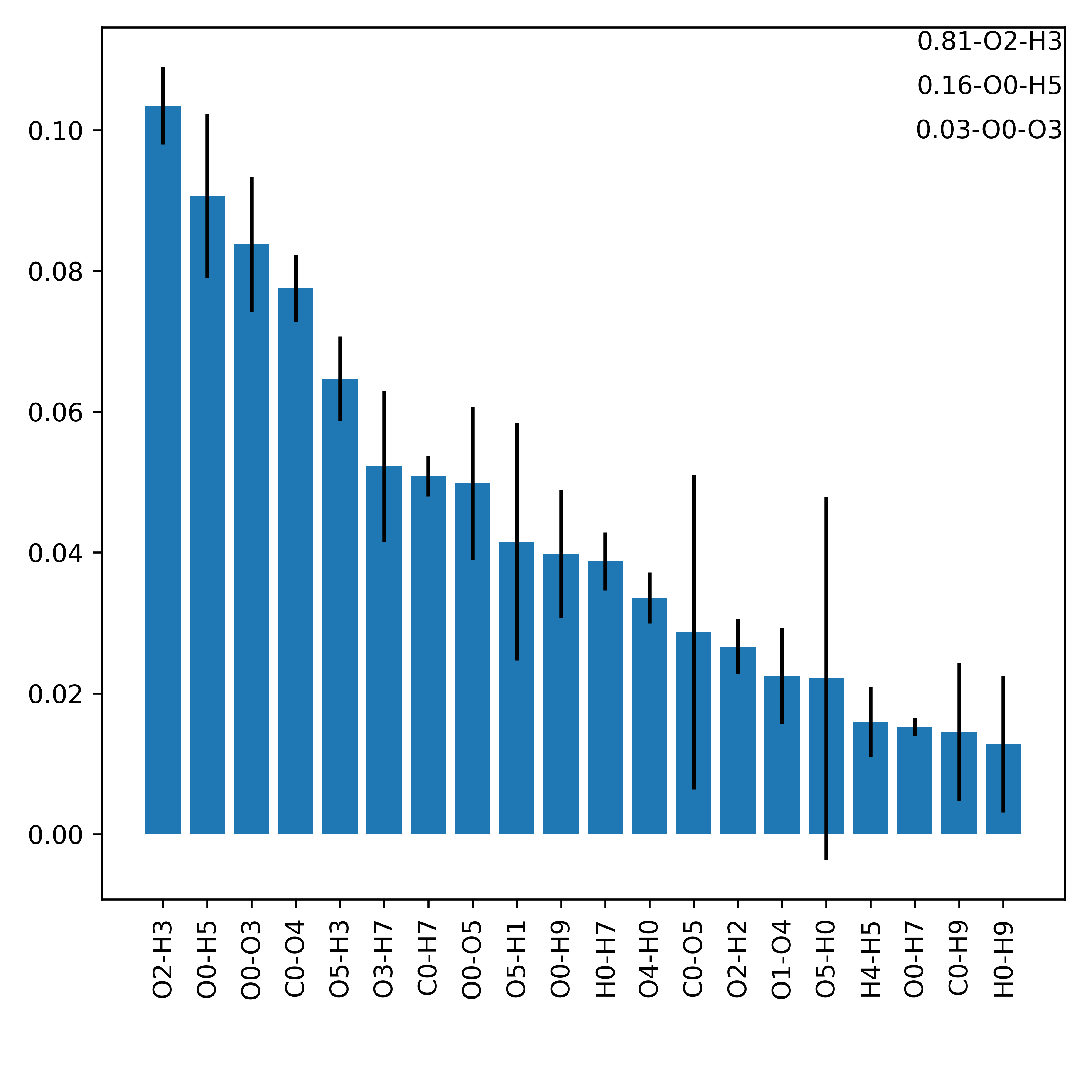}
    \caption{The importance approximations of the first split question from a random forest with depth of one for the four water case system.
    The bars represent the feature importance of a Random Forest, with the error bar calculated with a block-error average based on the generated trajectories.}
    \label{fig:4w_errors_ok}
\end{figure}

Our analysis generated the decision tree reported in Figure \ref{fig:4w_tree_ok}A. To simplify the visualization of the main splits that lead to the highest reactive trajectories of the decision tree, in  Figure \ref{fig:4w_tree_ok}B we report an \textit{xyz} representation. The atoms in blue, yellow, and green are involved in the first, second, and third split, respectively.

The deprotonation reaction of FA appears to primarily require the distance between O0 and O3 to be smaller than 2.56 Å. In other words, the first water molecule that accepts the proton from FA has to be sufficiently close. When this condition occurs, the probability to obtain a reactive path rises from 5\% to 12\%.

From Figure~\ref{fig:4w_errors_ok} (bottom) the probability that this is the most important feature is 3\%. Another feature has a probability of 16\%.  This is the distance between O0 and H5, the hydrogen that is connected to O3.  Therefore, this condition is actually equivalent.

The next split, along the branch that leads to the highest reactive probability, is the requirement that the distance between C0 and O4 is smaller than 4.20 Å. Qualitatively, this means a water molecule has to be closer than two hydrogen bonds away from FA. If this second requirement is also satisfied, the probability of a reactive path is of 30\%.

Still along the branch towards the highest reactive probability, the distance between O5 and H0 being smaller than 5.17 Å represents the last split here considered. This indicates that a second water molecule has to be within two hydrogen bonds away. When this additional requirement is satisfied, the probability of a reactive path reaches 58\%.

\subsection{Formic Acid with 6 water molecules}

\begin{figure}
    \centering
    \textbf{A}
    \includegraphics[width=\textwidth]{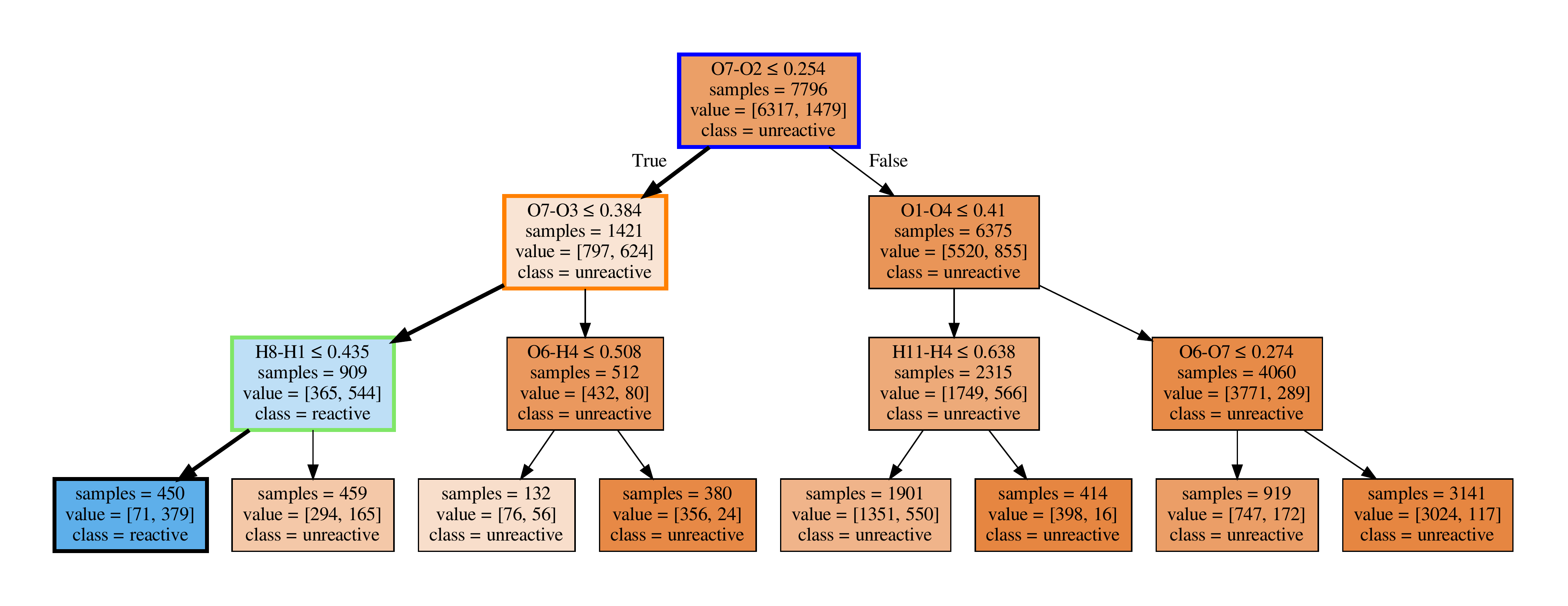}
    \\
    \textbf{B}
    \includegraphics[width=0.4\textwidth]{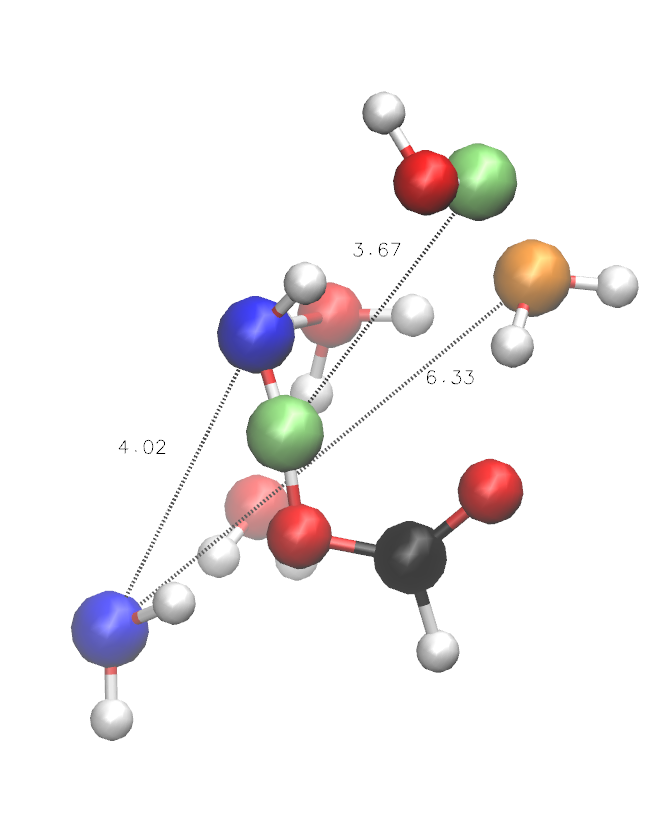}
    \caption{Decision tree for the system with 6 water molecules around the formic acid molecule. Three splits divide the input, each square reports 1) the question which splits the data, 2) the number of samples going into the node, 3) the number of [unreactive, reactive] samples going into the node and 4) the majority class of the node. At each split, the True branch is on the left, False on the right. The color indicates the ratio between unreactive (brown) and reactive (blue) samples included in a node. Wider arrows have been used to link the decision three split with the atoms involved. In panel B, a 3D representation of the system is provided. The atoms highlighted in blue, yellow, and green, correspond to the atoms involved in the first, second, and third split of the decision three, respectively. In red are the oxygen and in white the hydrogen atoms not indicated by the decision tree reported in the panel B.}
    \label{fig:6w_tree_ok}
\end{figure}

\begin{figure}
    \centering
    \includegraphics[width=0.92\textwidth]{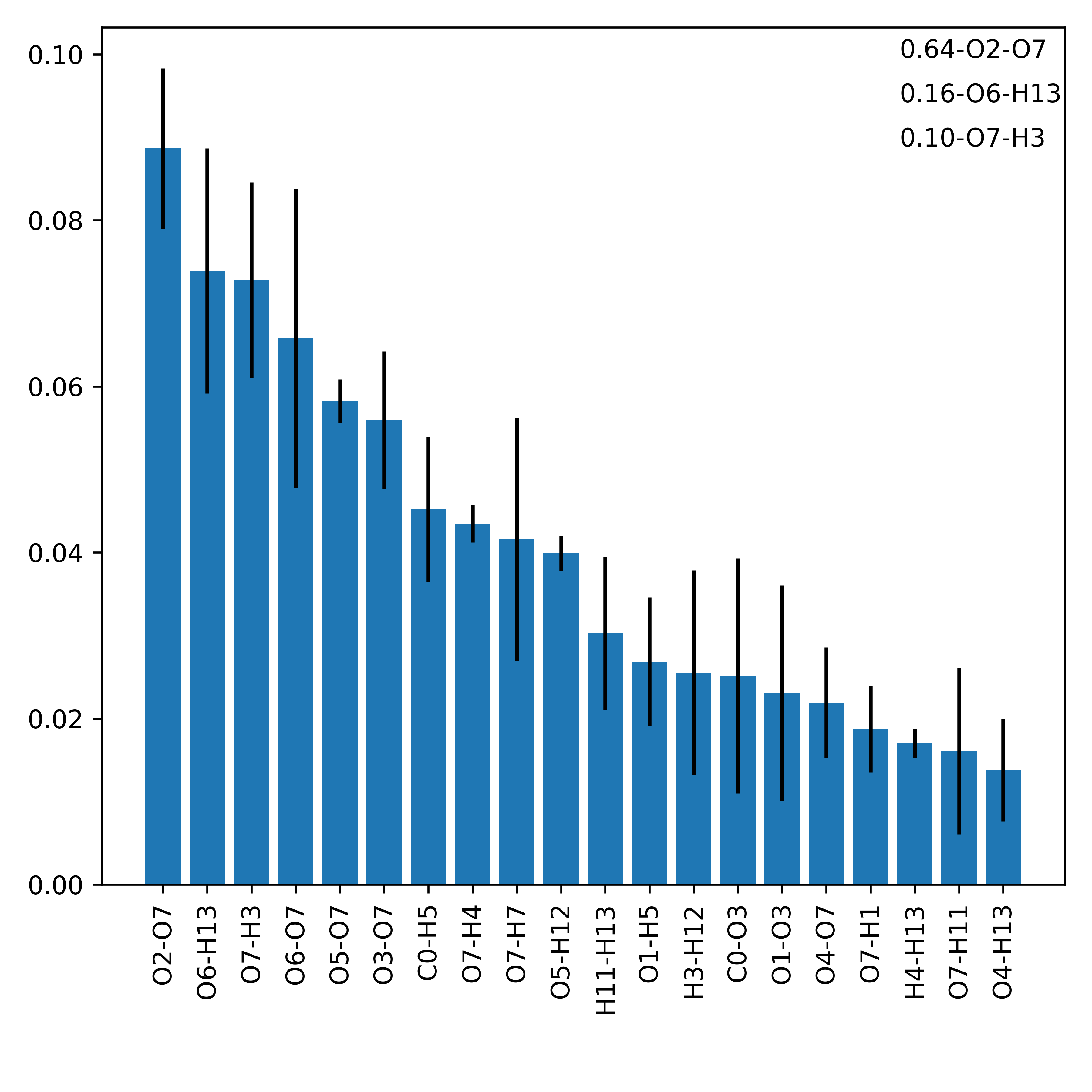}
    \caption{The importance approximations of the first split question from a random forest with depth of one for the four water systems.
    The bars represent the feature importance of a Random Forest, with the error bar calculated with a block-error average based on the generated trajectories.}
    \label{fig:6w_errors_ok}
\end{figure}

Figure~\ref{fig:6w_tree_ok} reports the analysis results obtained from the simulation of the proton transfer reaction for FA acid in a 6 water molecule cluster. In the figure, the decision tree and the \textit{xyz} structure which highlights the most probable splits which determine a reactive path are included.
 
In this system the deprotonation reaction of FA appears to primarily require the distance between O7 and O2 to be smaller than 2.54 Å. The first water molecule that accepts the proton from FA has to be sufficiently close. When this condition occurs, the probability to obtain a reactive path raises from 19\% to 44\%.

From Figure~\ref{fig:6w_errors_ok} (bottom) the probability that this is the most important feature is 3\%. One other equivalent feature has a probability of 16\%  which involves the distance between O0 and H5, the hydrogen that is connected to O3.

The next split, still along the branch that led to the highest reactive stance, is the distance between O7 and O3 being smaller than 3.84 Å. If a second water molecule is sufficiently close to the FA oxygen, the probability of a reactive path is 60\%.

Still along the branch towards the highest reactive stance, the distance between the hydrogen H8 and H1 being smaller than 4.35 Å represents the last split here considered. This feature can be interpreted as a structural requirement for the  water complex to be reactive, and may also implicitly involve some requirement involving atomic orientations/angles. In such conditions, the probability for the path to be reactive is 84\%.

The result reported by the decision trees generated by the index variant and index invariant representations are physically consistent but different in atom selection. That is, different atom pairs for the different representations are identified as most relevant. Yet, the water structure and formic acid orientation are equivalent. It should also be noted that the index variant representation results in a numerically more stable generation of random forests, as the importance values for the symmetric features are summed together. For the distribution of importance of the random forest reported in Figures \ref{fig:4w_errors_ok} and \ref{fig:6w_errors_ok}, the first splits are thus identified with a higher probability.

\bibliography{acs-achemso}